\documentclass[aps,prb,amsmath,amssymb,superscriptaddress,floatfix,showpacs,twocolumn]{revtex4-1}

\usepackage{graphicx}
\usepackage{mathdots}
\usepackage{epstopdf} 
\usepackage[usenames,dvipsnames]{color}
\usepackage{bm}

\usepackage{inputenc}
\usepackage{xcolor}
\usepackage{tikz}

\newcommand{\bra}[1]{\langle #1 |}
\newcommand{\ket}[1]{|  #1 \rangle}

\usepackage{booktabs}
\newcommand{\bean}{\begin{eqnarray}}
\newcommand{\eean}{\end{eqnarray}}
\usepackage[colorlinks,bookmarks=false,citecolor=blue,linkcolor=red,urlcolor=blue]{hyperref}

\begin{document}

\usetikzlibrary{matrix}
\usetikzlibrary{calc,fit}
\tikzset{%
  highlight1/.style={rectangle,color=blue!,fill=blue!15,draw,fill opacity=0.3,thick,inner sep=0pt}
}
\tikzset{%
  highlight2/.style={rectangle,color=gray!,fill=gray!15,draw,fill opacity=0.3,thick,inner sep=0pt}
}

\title{Subharmonic transitions and 
Bloch-Siegert shift in electrically driven spin resonance}

\author{Judit Romh\'anyi}
\affiliation{Leibniz-Institute for Solid State and Materials Research, IFW-Dresden, D-01171 Dresden, Germany}
\affiliation{Institute of Physics, E\"otv\"os University,
Budapest, Hungary}

\author{Guido Burkard}
\affiliation{Department of Physics, University of Konstanz, D-78457 Konstanz, Germany}

\author{Andr\'as P\'alyi}
\affiliation{Institute of Physics, E\"otv\"os University,
Budapest, Hungary}
\affiliation{MTA-BME Condensed Matter Research Group,
Budapest University of Technology and Economics, Budapest, Hungary}

\date{\today}

\begin{abstract} 
We theoretically study coherent subharmonic (multi-photon)
transitions of a harmonically driven spin.
We consider two cases: 
magnetic resonance (MR) with a misaligned, i.e., non-transversal
driving field, and
electrically driven spin resonance (EDSR)
of an electron confined in a one-dimensional, parabolic
quantum dot, subject to Rashba spin-orbit interaction.
In the EDSR case, we focus on the limit where the orbital level spacing of the
quantum dot is the greatest energy scale.
Then, we apply time-dependent Schrieffer-Wolff perturbation theory
to derive a time-dependent effective two-level Hamiltonian,
allowing to describe both MR and EDSR using the 
Floquet theory of periodically driven two-level systems. 
In particular, we characterise the fundamental (single-photon)
and the half-harmonic (two-photon)
spin transitions.
We demonstrate the appearance of two-photon Rabi oscillations,
and analytically calculate the fundamental and 
half-harmonic resonance frequencies and the corresponding 
Rabi frequencies.
For EDSR, we find that both the fundamental and the half-harmonic
resonance frequency changes upon 
increasing the strength of the 
driving electric field, which is an effect analogous to the 
Bloch-Siegert shift known from MR.
Remarkably, the drive-strength dependent correction to the 
fundamental EDSR resonance frequency has an anomalous, negative sign, 
in contrast to the corresponding Bloch-Siegert shift in MR which is always positive. 
Our analytical results are supported by numerical simulations,
as well as by qualitative interpretations for simple limiting cases. 
\end{abstract}

\pacs{
71.70.Ej 		
73.21.La 		
76.20.+q 		
}

\maketitle

\section{Introduction}
\label{sec:introduction}

Magnetic resonance (MR) is an established method to coherently control
the quantum state of spins. 
A simple example is a spin-$1/2$ electron 
subject to a time-dependent magnetic field 
\cite{Bloch-Siegert,Shirley,Koppens-esr,Veldhorst}, described
by the Hamiltonian 
\bean
\label{eq:esr}
H = -\frac 1 2 g\mu_B \mathbf{B}(t) \cdot \boldsymbol{\sigma},
\eean
where the magnetic field $\mathbf{B}(t) = (0,-B_{\rm ac} \cos\omega t, B)$
consists of a  `longitudinal' static component $B$ 
and a `transverse' ac component characterised by 
the drive strength $B_{\rm ac}$ and the drive frequency $\omega$,
and couples to the electron spin represented by the 
vector $\boldsymbol \sigma = (\sigma_x,\sigma_y,\sigma_z)$ 
of Pauli matrices. 

A typical initial-value problem considered in MR is
when  the initial state of the spin $\psi(t=0) = \ket{\uparrow}$ is the ground state
 of the static
Hamiltonian, $-\frac 1 2 g \mu_B B \sigma_z$,
and driving is switched on abruptly at $t=0$.
In the case of weak driving $B_{\rm ac} \ll B$, 
the rotating wave approximation (RWA) often provides a satisfactory description 
of the dynamics.
Using this approximation, one finds the following
simple phenomenology. 
If the resonance condition $\hbar \omega = g \mu_B B$ is 
fulfilled,
the drive will induce complete Rabi oscillations 
resulting in a transition probability 
$P_\downarrow(t) \equiv |\bra{\downarrow} \psi(t) \rangle|^2 = 
\sin^2( \Omega t/2)$,
where $\Omega =g \mu_B B_{\rm ac}/(2\hbar)$ 
is called the  Rabi frequency. 
Otherwise, i.e., in the case of a finite
detuning $\delta = \omega - g\mu_B B/\hbar$ between
the drive frequency and the resonance frequency, 
one finds incomplete Rabi oscillations with 
a $\delta$-dependent frequency: 
$P_\downarrow(t) = 
P^{\rm max}_\downarrow \sin^2(\sqrt{\delta^2 + \Omega^2} t/2)$,
with 
$P^{\rm max}_\downarrow = 
\Omega^2/(\Omega^2+\delta^2)< 1$.

Still focusing on the weak-driving regime $B_\textrm{ac} \ll B$,
one can go beyond the RWA, e.g., by numerical simulations
or analytical techniques such as 
the Floquet perturbation theory\cite{Shirley}.
Then, a richer phenomenology is revealed, including
(i) subharmonic or `multi-photon' resonances\cite{Shirley},
(ii) drive-strength-dependent Bloch-Siegert shifts\cite{Bloch-Siegert} (BSSs) 
of the resonance frequencies,
and 
(iii) Bloch-Siegert oscillations modulating the simple Rabi oscillations\cite{Bloch-Siegert}. 
We restrict our attention to (i) and (ii) here. 

(i) In the case of a transverse ac field,
such as the example used in Eq. \eqref{eq:esr},
odd subharmonic resonances appear\cite{Shirley}.
Rabi oscillations are obtained not only for
the fundamental resonance
$\omega \approx g\mu_B B/\hbar$, but also 
when $\omega \approx g\mu_B B/(N\hbar)$ with
$N=3,5,7,\dots$. 
In the case of a misaligned, non-transversal, ac field, 
such as 
$\mathbf B(t) = (0,-B_{\rm ac} \cos\theta \cos \omega t, B -B_{\rm ac} \sin \theta \cos \omega t)$ with $0< \theta < \pi/2$, 
both even and odd subharmonics are present. 
The Rabi frequency $\Omega^{(N)}_\textrm{res}$ 
at the $N$-photon subharmonic resonance is weaker than 
that of the fundamental one: $\Omega^{(N)}_\textrm{res} \propto B_{\rm ac}^N/B^{(N-1)}$.

(ii) The resonance frequencies $\omega_{\rm res}^{(N)}$ 
(i.e., the drive frequencies where complete Rabi oscillations
are induced) increase with increasing drive strength, 
by an amount that depends on $N$, and
is proportional to $B_{\rm ac}^2/B$. 

In many situations, it is be more convenient to control spins
using an ac electric field rather than an ac magnetic field. 
For example, if an electron spin is electrostatically confined
in a quantum dot (QD), then an ac electric field can be 
easily created by
applying an ac voltage component of the confinement
gate electrodes.
Along these lines, electrically driven 
spin resonance\cite{Golovach-edsr,Flindt,Tokura,Rashba-prb-edsr,Walls,Khomitsky,RuiLi} 
(EDSR) of
individual electron spins was demonstrated in a variety of
materials\cite{Kato-gfactorresonance,Nowack-esr,PioroLadriere-esr,Laird-prl,NadjPerge-spinorbitqubit,NadjPerge,FeiPei,Pribiag,Laird-nn,Kawakami-edsr,Klimov-edsr,Forster}.
As the ac electric field couples to the orbital degree of freedom
of the electron and has no direct effect on the spin, a sufficiently strong
coupling mechanism between the orbit and spin is required for EDSR.
Such a coupling can be supplied by spin-orbit interaction,
hyperfine interaction, or an inhomogeneous magnetic field. 

Recent experimental\cite{Laird-sst,Schroer-gfactor,Laird-nn,Stehlik-harmonic,Forster}
and theoretical\cite{De,Pingenot,Rashba-subharmonic-edsr,Nowak-harmonic,Osika-nanowire,Szechenyi-maximalrabi,Osika-nanotube,DanonRudner} 
studies addressed subharmonic resonances in EDSR.
One mechanism that leads to subharmonic resonances in EDSR 
is the appearance of higher harmonics $N\omega$ of the drive frequency
$\omega$ in the induced orbital 
dynamics\cite{Rashba-subharmonic-edsr,Osika-nanowire}. 
In this case, the time-dependent effective magnetic fields 
caused by the orbital dynamics will also have  
components at frequency $N\omega$,  leading to 
Rabi oscillations as $N  \omega$ matches
the Zeeman splitting.
Higher harmonics in the orbital dynamics arise naturally if the
confinement potential is anharmonic, or
if the driving electric field is inhomogeneous.
Subharmonic EDSR resonances can also arise in the presence
of harmonic confinement and homogeneous ac electric field, 
if the gradient of the effective magnetic fields is inhomogeneous;
this is the case, e.g., if the the effective magnetic field
is spatially localized or disordered\cite{Szechenyi-maximalrabi}.
A third mechanism, able to cause strong subharmonics with large $N$, 
is provided by Landau-Zener dynamics in the vicinity of 
level anticrossings\cite{Stehlik-harmonic,DanonRudner}. 

\begin{figure}
\includegraphics[width=0.9\columnwidth]{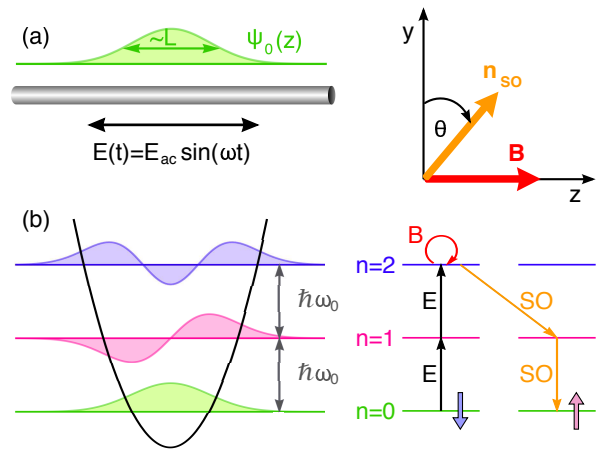}
\caption{
Electrically driven spin resonance in a 1D quantum dot. 
(a) An electron occupying the ground state $\psi_0(z)$ of a parabolic 
confinement potential  is excited by an ac electric field of
amplitude $E_{\rm ac}$ and frequency $\omega$. 
The electron is subject to a homogeneous magnetic field $\mathbf B$
and spin-orbit interaction characterised by the direction $\mathbf n_{\rm SO}$. 
(b) Left: Orbital levels labelled by the 
oscillator quantum number $n$,
separated by the level spacing $\hbar \omega_0$.
Right: Diagram representing one 
of the many fifth-order virtual processes contributing to 
the half-harmonic resonance. 
Horizontal lines represent the energy eigenstates of the
harmonic oscillator Hamiltonian.
Arrows labelled by $E$, $B$ and SO correspond to matrix elements of
the ac electric field, magnetic field and spin-orbit Hamiltonians, respectively.
Note that the spin-orbit Hamiltonian provides both spin-flip and 
spin-conserving matrix elements only if 
both $\cos \theta$ and $\sin \theta$ are nonzero.
[Cf. Eq. \eqref{eq:model}.]
}
\label{fig:setup}
\end{figure}

In this work, 
we theoretically
describe the characteristics of the half-harmonic resonance
in MR with a misaligned ac field, 
as well as in EDSR.
First, we use Floquet perturbation theory\cite{Shirley} 
to characterise the parameter dependence of the 
half-harmonic resonance frequency and the corresponding
Rabi frequency in the case of MR; in particular, 
the BSS is calculated. 
As for EDSR, we study a model\cite{Trif}
(see Fig. \ref{fig:setup}), where a
single electron is parabolically confined in 
a one-dimensional (1D) quantum dot, 
and is subject to a
dc magnetic field, an ac electric field, and 
spin-orbit interaction of Rashba type, the latter three being 
spatially homogeneous. 
We show that the half-harmonic resonance does arise in this model, 
despite the harmonic confinement and homogeneous ac electric field.
In the perturbative regime of this model, i.e., when 
the orbital level spacing $\hbar \omega_0$ 
dominates over other energy scales,
we analytically derive the parameter dependence of the
half-harmonic resonance frequency and 
the corresponding Rabi frequency. 
This is achieved via a combination of time-dependent Schrieffer-Wolff
perturbation theory (TDSW), which is used to obtain a $2\times 2$ effective 
`two-level' or `qubit' Hamiltonian
$\tilde{\mathcal H}_q$, 
and Floquet perturbation theory, applied to describe the qubit dynamics 
governed by $\tilde{\mathcal H}_q$. 

This paper is structured as follows. In Sec.~\ref{sec:summaryofresults} we summarize our main results. Secs.~\ref{sec:esr} and~\ref{sec:edsr} are dedicated to the detailed discussion of MR and EDSR, respectively. In them we formulate the problems, derive analytical solutions and compare these with numerical simulations where called for. In Sec.~\ref{sec:conclusions} we give a conclusion of our findings. 

\section{Summary of the results}
\label{sec:summaryofresults}

In this Section, we summarize the main results that are derived in later
Sections.
Let us start with the case of spin-1/2 MR with a misaligned ac field.
The parameters of the model are $\tilde B$,
the energy scale of the static magnetic field along the
z axis;
$\tilde B_\textrm{ac}$, 
the energy scale of the ac magnetic field oriented in the yz plane;
and $\theta$, the angle enclosed by the ac field and the y axis. 
(For more details, see Sec. \ref{sec:esr}.)
The fundamental (or single-photon) resonance frequency 
$\omega^{(1)}_\textrm{res}$, 
that is, the drive frequency at which the Rabi oscillations are complete, 
deviates from the Zeeman splitting:
\bean
\label{eq:mrfhres}
\hbar \omega^{(1)}_\textrm{res} = \tilde B + \hbar \omega^{(1)}_\textrm{BSS},
\eean
where the second term on the right hand side is the 
BSS and has the form
\bean
\label{eq:mrbss1}
\hbar \omega^{(1)}_\textrm{BSS} = 
\frac{\tilde B_\textrm{ac}^2 \cos^2\theta}{16 \tilde B}.
\eean
The Rabi frequency at the fundamental resonance is
\bean
\label{eq:mrfhrabi}
\hbar \Omega^{(1)}_\textrm{res} = \frac{\tilde B_\textrm{ac}}{2} \cos \theta.
\eean
Similarly to the fundamental resonance, the 
half-harmonic (two-photon) resonance also 
acquires a positive BSS: 
\bean
\label{eq:mrhhres}
\hbar \omega^{(2)}_\textrm{res} = \frac{\tilde B}{2} + \hbar \omega^{(2)}_\textrm{BSS},
\eean 
where 
\bean
\label{eq:mrbss2}
\hbar \omega^{(2)}_\textrm{BSS} = 
\frac{\tilde B_{\textrm{ac}}^2\cos^2\theta}{6\tilde B}.
\eean
The Rabi frequency at the half-harmonic resonance is
\bean
\label{eq:mrhhrabi}
\hbar \Omega^{(2)}_\textrm{res} = \frac{\tilde B_\textrm{ac}^2 \sin 2 \theta}{4\tilde B}.
\eean
Note that the resonance frequencies above
are expressed up to second order in the small energy scale 
$\tilde B_\textrm{ac} \ll \tilde B$.  
For a detailed discussion of these results, see Sec. \ref{sec:mrdiscussion}.

In the case of EDSR in a 1D parabolic QD, the parameters characterizing the model are 
as follows. 
The orbital level spacing $\hbar \omega_0$ is the dominant energy scale; 
the static magnetic field, oriented along the z axis, 
is characterized by the Zeeman-splitting 
energy scale $\tilde B$; 
spin-orbit interaction is described by the energy scale $\tilde \alpha$ and
the unit vector ${\bf n}_\textrm{so}=(0,\cos\theta,\sin \theta)$ which points along
the spin-orbit field;
and the energy scale $\tilde E_\textrm{ac}$ describing the strength of the
driving ac electric field.
We find the following results for the fundamental resonance frequency:
\bean
\label{eq:fund_res}
\hbar \omega^{(1)}_\textrm{res} = 
\tilde B + \hbar \omega^{(1)}_{g} + \hbar \omega^{(1)}_\textrm{nlZ}
+ \hbar \omega^{(1)}_\textrm{BSS},
\eean
where the correction consists of a g-factor renormalization term,
\bean
\hbar \omega^{(1)}_{g} = 
-\frac{2\tilde B\tilde\alpha^2\cos^2\theta}{\hbar^2\omega^2_0}\left(1-\frac{\tilde\alpha^2(1+\sin^2\theta)}{\hbar^2\omega^2_0}\right)\;,
\eean
a term describing the non-linear Zeeman effect,
\begin{equation}
\hbar \omega^{(1)}_\textrm{nlZ} =
\frac{2 \tilde B^3\tilde\alpha^2\cos^2\theta}{\hbar^4\omega^4_0}\;,
\end{equation}
and a correction that is second order in the 
drive strength $\tilde E_\textrm{ac}$, hence
analogous to the BSS:
\begin{equation}
\hbar \omega^{(1)}_\textrm{BSS} = -
\frac{\tilde B \tilde\alpha^2\tilde E^2_\textrm{ac}\cos^2\theta}{\hbar^4\omega^4_0}\;.
\end{equation}
Note that the sign of this BSS is negative, in contrast to the 
positive sign in the case of
MR [Eqs. \eqref{eq:mrbss1} and \eqref{eq:mrbss2}];
an interpretation of this anomalous sign is given in Sec. \ref{sec:conclusions},
paragraph (4).
The Rabi frequency at the fundamental resonance is
\bean
\label{eq:Rabi_at_fund_res}\hbar \Omega^{(1)}_\textrm{res}&=&2\frac{\tilde B\tilde E_{\sf ac}\tilde\alpha\cos\theta}{\hbar^2\omega^2_0}\left(1+\frac{\tilde B^2-2\tilde\alpha^2}{\hbar^2\omega^2_0}\right).
\eean
The half-harmonic resonance frequency is shifted with respect to the
half of the fundamental resonance frequency:
\bean
\label{eq:halfharmonicresonancefrequency}
\hbar \omega_\textrm{res}^{(2)}& = &
\frac 1 2 \left(
	\tilde B 
	+ \hbar \omega_{g}^{(1)}
	+ \hbar \omega_\textrm{nlZ}^{(1)}
\right)
+
\hbar \omega^{(2)}_\textrm{BSS},
\eean
where the drive-strength-dependent
BSS is expressed as
\bean
\hbar \omega^{(2)}_\textrm{BSS} =
\frac{2 \tilde B \tilde \alpha ^2 \tilde E^2_\textrm{ac} \cos^2\theta  }{3 \hbar^4 \omega_0^4}.
\eean
The Rabi frequency at the half-harmonic resonance is
\bean
\label{eq:RabiAtHalfHarmonic}
\hbar \Omega_\textrm{res}^{(2)} &=&
\frac{\tilde B \tilde \alpha ^2 \tilde E_{\rm ac}^2 \sin(2\theta)}{ (\hbar \omega_0)^4}.\eean
Note that the EDSR resonance and Rabi frequencies above
are expressed up to fifth order in the small energy scales
$\tilde \alpha, \tilde B, \tilde E_\textrm{ac} \ll \hbar \omega_0$. 
A detailed discussion of these results, and their comparison with numerical
solutions of the time-dependent Schr\"odinger equation, 
is included in Sec. \ref{sec:results}.

\section{Magnetic resonance with a misaligned ac field}
\label{sec:esr}

In this Section, using Floquet perturbation theory, 
we derive and discuss the properties of the fundamental (single-photon)
and half-harmonic (two-photon) resonances
in MR, for the spin-1/2 case.
In particular, the results \eqref{eq:mrfhres}---\eqref{eq:mrhhrabi} are derived.

\subsection{Problem formulation}

We consider MR spin dynamics 
driven by a misaligned ac field. 
The Hamiltonian reads
\begin{eqnarray}
\mathcal H(t) = -\frac{1}{2}\tilde{\bf B}(t) \cdot \boldsymbol{\sigma}\;,
\label{eq:ham_esr}
\end{eqnarray}
where the magnetic field has the form 
\bean
\label{eq:mrfield}
\tilde{\bf B}(t) =
\left(\begin{array}{c}
0 \\
-\tilde{B}_\textrm{ac} \cos\theta \cos \omega t \\
\tilde{B} - \tilde{B}_\textrm{ac}\sin \theta \cos \omega t
\end{array} \right).
\eean
Here we introduced 
$\tilde B = g \mu_B B$ and $\tilde B_\textrm{ac} = g\mu_B B_\textrm{ac}$.
Henceforth the parameters with tilde, e.g., $\tilde B$,
have energy dimension.
The Hamiltonian in Eq. \eqref{eq:ham_esr} has four parameters: 
the strength of the static field $\tilde B$, 
the strength of the driving field $\tilde B_\textrm{ac}$, the
frequency of the driving field $\omega$,
and the misalignment angle $\theta$.
Note that $\theta = 0$ corresponds to a transverse ac field, 
and $\theta = \pi/2 $ corresponds to a longitudinal ac field. 
We consider the case of weak driving, $\tilde B_\textrm{ac} \ll \tilde B$. 

In particular, we want to solve the initial-value
problem described in Sec. \ref{sec:introduction}:
the initial state is the ground state $\ket{\uparrow}$ of the
 Hamiltonian without driving i.e., $\psi(t=0) = \ket{\uparrow}$, 
driving is switched on abruptly at $t=0$, 
and we are interested in the time evolution $\psi(t)$ of this state.
We calculate the \emph{transition probability} describing
the time-dependent occupation of the excited state $\ket{\downarrow}$
at the fundamental and half-harmonic resonances, 
and from those we deduce the parameter dependencies of the resonance frequencies
and the Rabi frequencies. 

In the rest of this Section, we use Floquet perturbation 
theory\cite{Dittrich,Shirley,Aravind1984,Gromov}
to derive the results and to provide qualitative interpretations
in simple limiting cases, such as the limits of transversal and longitudinal
ac fields.
Even though similar treatments can be found in the 
literature\cite{Shirley,Gromov}, we present a detailed 
discussion of the MR problem for 
the following reason.
The MR problem is relatively simple as compared to the
EDSR problem, which can be appreciated, e.g., by comparing the 
 driven two-level Hamiltonians of Eqs. \eqref{eq:ham_esr} and \eqref{eq:edsreffectiveh}, respectively. 
Moreover, as we will show, the Floquet method  and the qualitative interpretations
we describe here for the MR problem
can be carried over to the EDSR problem, once the $2\times 2$ 
effective qubit Hamiltonian in Eq. \eqref{eq:edsreffectiveh} is obtained for the latter. 
This allows us to provide a rather compact description of
the EDSR in the forthcoming Sections, 
by referencing this Section wherever possible.

\subsection{Floquet method}
\label{sec:floquet}

The Floquet method allows one to find the solution 
of an initial-value problem of a periodically driven quantum system,
described by the time-periodic Hamiltonian $H(t) = H(t+T)$. 
The period of the driving is denoted by $T$, and the corresponding
(angular) frequency by $\omega = 2\pi / T$. 
The key ingredient of the method is the quantum-mechanical Floquet
theorem\cite{Dittrich}, which guarantees that
the Schr\"odinger equation $i\hbar\dot\Psi(t)=\mathcal{H}(t)\Psi(t)$
of a $d$-level system has  $d$ solutions $\Psi_k(t)$ ($k=1,\dots,d$) 
that are themselves periodic with period $T$, apart from a
phase factor. 
Therefore, these special solutions have the form
\begin{eqnarray}
\label{eq:stationarystates}
|\Psi_k(t)\rangle
=
e^{-i E_k t/\hbar}\sum_{l=1}^{d} \sum^{\infty}_{m=-\infty}
c_{k,lm}e^{i m \omega t}|\psi_{l}\rangle\;,
\end{eqnarray}
where $\ket{\psi_l}$ is an arbitrary basis of the Hilbert space. 
Note that the result of the double sum is a periodic function of $t$ with period $T$. 
In Eq. \eqref{eq:stationarystates}, the quantity 
$E_k$ and the coefficients $c_{k,lm}$ are a priori unknown;
the former is called \emph{quasi-energy}. 
Once these special solutions $\ket{\Psi_k(t)}$ are found, 
they provide the propagator 
\bean
\label{eq:floquetpropagator}
U(t,0) = \sum_{k=1}^d \ket{\Psi_k(t)} \bra{\Psi_k(0)},
\eean
which in turn provides the solution of any initial-value problem 
via 
\bean
\label{eq:initialvalue}
\Psi(t) = U(t,0) \Psi(0). 
\eean

The special solutions $\Psi_k(t)$ are found
by using Eq. \eqref{eq:stationarystates} as an Ansatz,
substituting it to the Schr\"odinger equation, 
evaluating the scalar product of the equation
with $\bra{\psi_{l'}}$, 
multiplying the equation by $e^{-im' \omega t}$ and
integrating the equation in time between $t=0$ and $t=T$. 
This procedure yields the following eigenvalue equation for $E_k$:
\bean
\sum_{l=1}^d \sum_{m = -\infty}^{\infty}
\mathcal F_{l' m', lm} c_{k,lm} = E_k c_{k,l'm'},
\label{eq:floquetschrodinger}
\eean
where 
\bean
\mathcal F_{\!l'm',lm\!}\! =\! m \hbar \omega \delta_{l' l} \delta_{m' m}\!+\!
\!\sum_{n=-\infty}^\infty\! 
\bra{\psi_{l'}} \mathcal H^{\!(n)\!} \ket{\psi_l}
\delta_{\!m'\!,n\!+\!m}
\eean
is the \emph{Floquet matrix} or \emph{Floquet Hamiltonian},
and we introduced the Fourier components
$\mathcal H^{(n)}$ of the Hamiltonian via
\bean
\mathcal H(t) = \sum_{n=-\infty}^\infty \mathcal H^{(n)} e^{in\omega t}.
\eean

We call two eigenvalue-eigenvector pairs  of $\mathcal F$ equivalent,
if the two time-dependent solutions they generate via
Eq. \eqref{eq:stationarystates} are the same.
Importantly, even though the number of eigenvalue-eigenvector pairs
of $\mathcal F$ is infinite, they form only 
 $d$ equivalence classes.

In summary, we have transformed the 
time-dependent Schr\"odinger equation of the 
periodically driven $d \times d$ Hamiltonian $\mathcal H(t)$ into 
the time-independent Schr\"odinger equation \eqref{eq:floquetschrodinger}
of the infinite-dimensional Floquet Hamiltonian $\mathcal F$.
To construct the special solutions \eqref{eq:stationarystates},
and thereby the solution of any initial-value problem
via Eqs. \eqref{eq:floquetpropagator} and \eqref{eq:initialvalue},
the  quasi-energies $E_k$ and
the corresponding eigenvectors $c_k$ 
should be found by solving the eigenvalue problem of 
the Floquet Hamiltonian $\mathcal F$.

\subsection{Perturbative description of the transition probability}

\label{sec:mrfloquetperturbation}

After reviewing the Floquet method in general, 
we now apply this to the MR problem defined in 
Eq. \eqref{eq:ham_esr}.
Here, we have a two--level system, therefore $d=2$, and we use
$\ket{\alpha} \equiv \ket{\uparrow} \equiv \ket{\psi_1} $ 
and 
$\ket{\beta} \equiv\ket{\downarrow} \equiv  \ket{\psi_2}$ to denote these levels. The Fourier components of the Hamiltonian read
\begin{subequations}
\bean
\mathcal H^{(0)} &=& - \frac 1 2 \tilde B \sigma_z, \\
\mathcal H^{(\pm 1)} &=& \frac 1 4 \tilde B_\textrm{ac} 
(\cos \theta \sigma_y + \sin \theta \sigma_z),
\eean
\end{subequations}
and the other Fourier components are zero.
\begin{figure}[b]
\includegraphics[width=0.95\columnwidth]{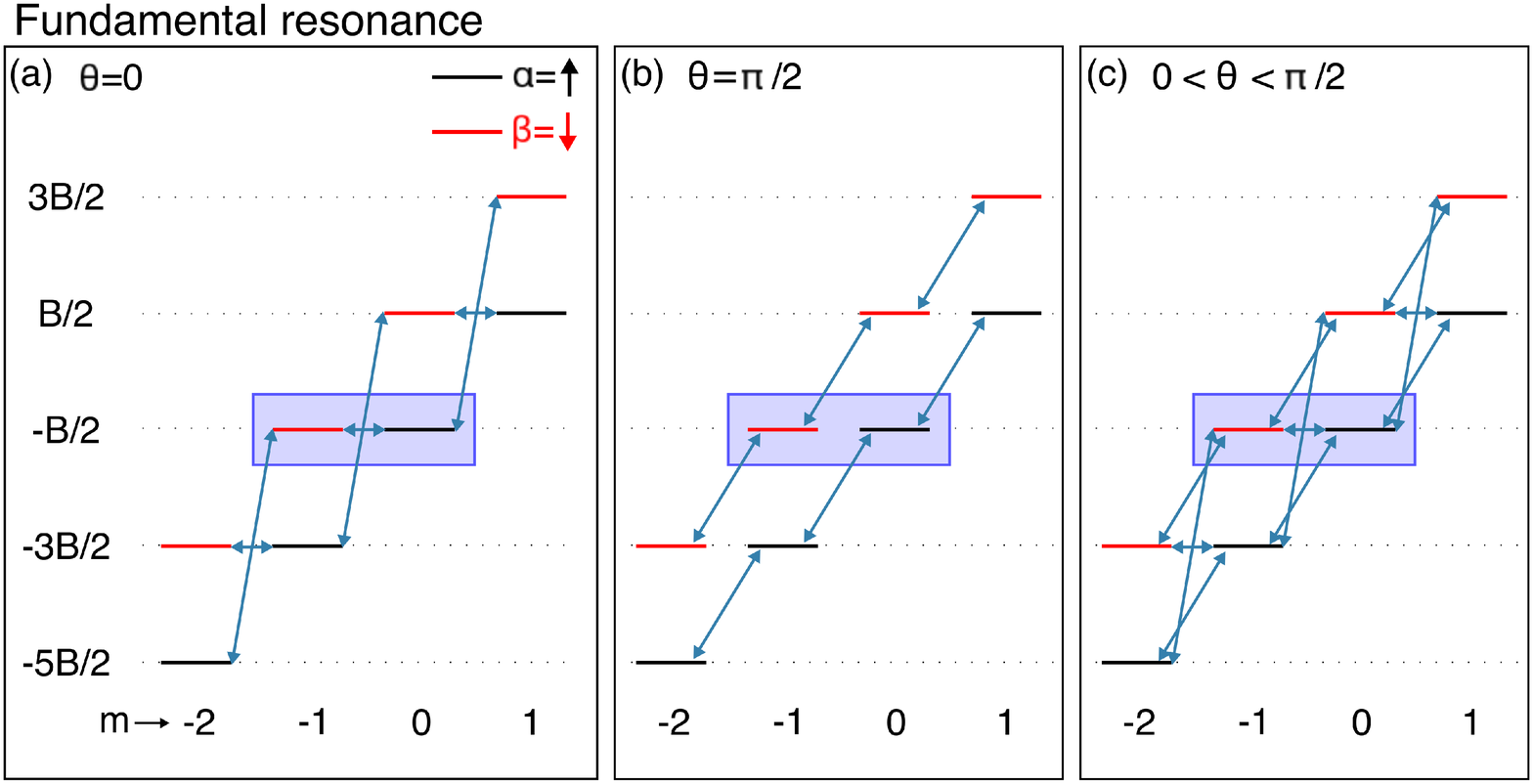}
\caption{
Magnetic resonance in a misaligned ac field: 
structure of the Floquet Hamiltonian at the fundamental resonance. 
 Panels show cases when the ac field
 is perpendicular to the static field (a),
 is parallel to the static field (b),
 has finite perpendicular and parallel components (c).
 Horizontal lines (blue arrows) correspond to diagonal (off-diagonal) 
 matrix elements of the  Floquet Hamiltonian $\mathcal{F}$.
 The vertical position of each horizontal line corresponds to
 the value of the diagonal matrix element. 
}
\label{fig:mrfundamental}
\end{figure}

First, consider the case when the drive frequency is close
to the fundamental resonance, $\hbar \omega \approx \tilde B$.
Then, the diagonal elements of the 
Floquet Hamiltonian $\mathcal F$ 
(\emph{Floquet levels}) form pairs:
\begin{equation} 
\mathcal F_{\alpha m, \alpha m}
= m \hbar \omega - \frac 1 2 \tilde B \approx
(m-1) \hbar \omega + \frac 1 2 \tilde B = \mathcal F_{\beta,m-1,\beta,m-1}.
\end{equation}
The distance between different pairs is 
approximately $ \hbar \omega \approx \tilde B$, 
which is much larger than the energy scale $\tilde B_\textrm{ac}$ 
characterising the off-diagonal elements of $\mathcal{F}$. 
Therefore, the tools of quantum-mechanical perturbation theory 
can be used to provide an approximate solution  of
the eigenvalue problem of the
Floquet Hamiltonian.

The structure of the Floquet Hamiltonian $\mathcal F$ 
is visualised for the case $\hbar \omega = \tilde B$ in 
the level diagram shown in Fig. \ref{fig:mrfundamental}.
Horizontal lines represent the diagonal matrix elements
$\mathcal F_{lm,lm}$
of the Floquet Hamiltonian, 
their vertical positions correspond to their
value, 
their colour (black, red) represents their spin index
$l \in (\alpha,\beta)$ and
their horizontal position stands for their
Floquet index $m = \dots, -1, 0, 1, 2, \dots$. 
The vertical spacing of the Floquet levels is 
$\hbar \omega = \tilde B$. 
The blue arrows indicate the nonzero off-diagonal matrix elements
of $\mathcal{F}$, which are of the order of $B_\textrm{ac}$ and
hence small compared to the level spacing. 

In the case $\hbar \omega = \tilde B$ shown in 
Fig. \ref{fig:mrfundamental}, 
the Floquet levels
form degenerate pairs. 
The  pair formed by 
$\mathcal F_{\beta,-1,\beta,-1}$
and 
$\mathcal F_{\alpha,0,\alpha,0}$ is 
highlighted in Fig. \ref{fig:mrfundamental} by the blue box.
The subspace of this pair is weakly coupled to the other
Floquet levels, hence this coupling can be treated perturbatively using 
(time-independent) Schrieffer-Wolff perturbation theory\cite{Schrieffer1966},
which is also known as quasi-degenerate perturbation theory\cite{Winkler}. 
This perturbative treatment is also applicable if
there is a finite, but small detuning
$\delta =  \omega - \tilde B/\hbar \ll \tilde B/\hbar $ from the resonance condition. 
The small dimensionless parameter characterising the
strength of the perturbation is $\epsilon = \tilde B_\textrm{ac}/ \tilde B$. 

In this case, the Floquet Hamiltonian reads

\begin{widetext}
\begin{equation}\mathcal{F}=
\begin{tikzpicture}[baseline=(m.center)]
  \filldraw[fill=blue!15, fill opacity=0.3,thick,draw=blue] (0.67,0.8) -- (-3.2,0.8) --(-3.2,-0.7)-- (0.67,-0.7)--(0.67,0.8);
   \filldraw[fill=gray!15, fill opacity=0.3,thick,draw=gray](-5.7,0.8) -- (-5.7,2.2)--(-3.2,2.2)--(-3.2,0.8)--(-5.7,0.8);
   \filldraw[fill=gray!15, fill opacity=0.3,thick,draw=gray](-5.7,-0.7) -- (-5.7,-3.7)--(-3.2,-3.7)--(-3.2,-0.7)--(-5.7,-0.7);
   \filldraw[fill=gray!15, fill opacity=0.3,thick,draw=gray](7.2,-0.7) -- (7.2,-3.7)--(0.67,-3.7)--(0.67,-0.7)--(7.2,-0.7);
   \filldraw[fill=gray!15, fill opacity=0.3,thick,draw=gray](7.2,0.8) -- (7.2,2.2)--(0.67,2.2)--(0.67,0.8)--(7.2,0.8);

    \matrix (m) [matrix of math nodes, left delimiter={.}, right delimiter={.},
     row sep=1mm, nodes={minimum width=2em, minimum height=1em}] {
{} & {} & {}  &  \alpha_{-1}&  \beta_{-1}  & \alpha_{0}  & \beta_{0}   & \alpha_{1}  & \beta_{1} & {} \\
{} & {} & {}  &  \downarrow &  \downarrow  & \downarrow  & \downarrow  & \downarrow  & \downarrow & {} \\
{}& {} & {}  &  \vdots &  \vdots  & \vdots  & \vdots  & \vdots & \vdots  & {} \\
 \alpha_{-1} & \rightarrow & \hdots & -\frac{1}{2}\tilde B-\hbar\omega & 0 & \frac{1}{4} \tilde B_\textrm{ac}\sin\theta & -\frac{i}{4} \tilde B_\textrm{ac}\cos\theta  &  0 & 0 & \hdots \\
 \beta_{-1} & \rightarrow & \hdots & 0  & \frac{1}{2}\tilde B- \hbar\omega & \frac{i}{4} \tilde B_\textrm{ac}\cos\theta& -\frac{1}{4} \tilde B_\textrm{ac}\sin\theta &  0  & 0 & \hdots\\
 \alpha_{0} & \rightarrow & \hdots & \frac{1}{4} \tilde B_\textrm{ac}\sin\theta  & -\frac{i}{4} \tilde B_\textrm{ac} \cos\theta & -\frac{1}{2}\tilde B &  0 & \frac{1}{4}  \tilde B_\textrm{ac} \sin\theta &-\frac{i}{4} \tilde B_\textrm{ac}\cos\theta  & \hdots\\
 \beta_{0} & \rightarrow &\hdots  & \frac{i}{4} \tilde B_\textrm{ac}\cos\theta & -\frac{1}{4} \tilde B_\textrm{ac}\sin\theta & 0 & \frac{1}{2}\tilde B&   \frac{i}{4} \tilde B_\textrm{ac}\cos\theta & -\frac{1}{4}  \tilde B_\textrm{ac} \sin\theta & \hdots \\
 \alpha_{1} & \rightarrow & \hdots  & 0 & 0 & \frac{1}{4}  \tilde B_\textrm{ac} \sin\theta & -\frac{i}{4} \tilde B_\textrm{ac}\cos\theta  &  -\frac{1}{2}\tilde B+ \hbar\omega & 0 & \hdots \\
 \beta_{1} & \rightarrow & \hdots & 0 & 0 & \frac{i}{4} \tilde B_\textrm{ac}\cos\theta & -\frac{1}{4} \tilde B_\textrm{ac} \sin\theta  & 0 &  \frac{1}{2}\tilde B+ \hbar\omega & \hdots \\
{} & {} & {} & \vdots & \vdots & \vdots & \vdots  & \vdots & \vdots  & {} \\
      }; 
\end{tikzpicture} 
\label{eq:Floq_esr}
\end{equation}
\end{widetext}

\subsubsection{Fundamental resonance within RWA}
\label{sec:mrfundamentalRWA}

Using first-order perturbation theory, the two non-equivalent
eigenvalues and eigenvectors of $\mathcal F$ can be found
approximately.
For this we introduce 
$\mathcal F_0$ and $\mathcal F_1$ so that
$\mathcal{F} = \mathcal{F}_0 + \mathcal{F}_1$.
 $\mathcal{F}_0$ is the diagonal component
of $\mathcal F$ at $\omega=\tilde B/\hbar$., i.e. at $\delta=0$.
%
 
First-order perturbation theory in $\mathcal F_1$ amounts to 
diagonalizing the $2\times 2$ block highlighted 
in purple in Eq. \eqref{eq:Floq_esr} and Fig. \ref{fig:mrfundamental}.
For future reference, we recast this $2\times 2$ block to
the form
\begin{eqnarray}
\tilde{\mathcal F} = 
\left[\begin{array}{cc}
\epsilon_0+\Delta & i \lambda\\
-i \lambda &\epsilon_0-\Delta
\end{array}
\right]\label{eq:gen2lev},
\end{eqnarray}
where
$\epsilon_0 =  -\frac 1 2 (\tilde B + \hbar \delta)$,
$\Delta = -\hbar \delta / 2$, and
$\lambda =  \frac 1 4 \tilde B_\textrm{ac} \cos \theta$.
The matrix $\tilde{\mathcal F}$ has eigenvalues
\begin{equation}
\label{eq:epm}
\tilde E_\pm=\epsilon_0\pm\sqrt{\Delta^2+\lambda^2}\;,
\end{equation}
and corresponding eigenvectors
\begin{equation}
\tilde c_\pm=N_{\pm}[\frac{i}{\lambda}(\Delta\pm\sqrt{\Delta^2+\lambda^2}),1]\;,
\label{eq:chi_pm}
\end{equation}
where $N_{\pm}$ is a normalization constant.
Note that instead of using the numerical index $k \in (1,2)$
labelling the solutions \eqref{eq:stationarystates}, 
in Eqs \eqref{eq:epm} and after
we use the values $k \in (+,-)$.

The results \eqref{eq:epm} and \eqref{eq:chi_pm} 
imply that the two non-equivalent approximate eigenvalue-eigenvector
pairs of $\mathcal F$ are
$(\tilde E_\pm,c_\pm)$, 
where 
\bean
c_{\pm,lm} = \left\{
\begin{array}{ll}
	\tilde c_{\pm,1} & \mbox{if $(l,m)=(\beta,-1)$,} \\
	\tilde c_{\pm,2} & \mbox{if $(l,m)=(\alpha,0)$,} \\
	0 & \mbox{otherwise.}
\end{array}
\right., 
\eean
and $\tilde c_{\pm,1}$ and $\tilde c_{\pm,2}$ are the components
of $\tilde c_{\pm}$ in Eq. \eqref{eq:chi_pm}.
This result allows us to construct the transition probability 
$P_{\beta \leftarrow \alpha}(t) = |\bra{\beta} \Psi(t) \rangle |^2$
from the initial spin (ground) state $\ket{\uparrow} \equiv \ket{\alpha}$
to the excited state $\ket{\downarrow} \equiv \ket{\beta}$
via Eqs. \eqref{eq:stationarystates}, \eqref{eq:floquetpropagator}
and \eqref{eq:initialvalue}.
A straightforward calculation yields
\bean
\label{eq:Pmrfundamental}
P_{\beta \leftarrow \alpha}(t) =
\frac{\lambda^2}{\lambda^2+\Delta^2} 
\sin^2\left(\frac{1}{\hbar}\sqrt{\Delta^2+\lambda^2} \, t \right).
\eean

According to Eq. \eqref{eq:Pmrfundamental},
the spin makes complete Rabi oscillations 
if $\Delta =0$, that is, $\delta = \omega - \tilde B/ \hbar= 0$.
Hence the single-photon resonance frequency is
$\omega_\textrm{res}^{(1)} = \tilde B/\hbar$.
The  Rabi frequency upon resonant driving is
$\Omega^{(1)}_\textrm{res} = 2\lambda/\hbar = \frac {\tilde B_\textrm{ac} }{2\hbar} \cos \theta$;
thus only the transverse component of the ac field 
contributes to the Rabi frequency at the fundamental resonance.
In fact, the result \eqref{eq:Pmrfundamental} is equivalent to the
one obtained by neglecting the longitudinal 
ac field and performing RWA.

\subsubsection{Fundamental resonance beyond the RWA: Bloch-Siegert shift of the resonance frequency}
\label{sec:mrfundamentalbeyondRWA}

Let us discuss the corrections to
$\omega_\textrm{res}^{(1)}$ and
$\Omega_\textrm{res}^{(1)}$ beyond the RWA. 
To this end, we incorporate in the analysis the effect of those matrix elements
of $\mathcal F_1$ that connect the two highlighted  Floquet
levels [see Eq. \eqref{eq:Floq_esr} and Fig. \eqref{fig:mrfundamental}] 
to the complementary subspace.
This is done via a (time-independent) Schrieffer-Wolff transformation 
that is second order in $\mathcal F_1$.
The resulting effective $2\times 2 $ Floquet Hamiltonian
$\tilde{\mathcal{F}}$ has the form given in 
Eq. \eqref{eq:gen2lev}, with 
\begin{subequations}
\bean
\label{eq:mrfundamentalDelta}
\Delta &=& 
-\frac{1}{2}\hbar \delta+\frac{\tilde B_\textrm{ac}^2\cos^2\theta}{32 \tilde B}\label{eq:delta_fn}
\\
\label{eq:lambda}
\lambda &=& \frac{\tilde B_\textrm{ac}\cos\theta}{4}.
\eean
\end{subequations}

Recall that the eigenvalues and eigenvectors of $\tilde{\mathcal F}$ 
are given by Eqs. \eqref{eq:epm} and \eqref{eq:chi_pm}.
From these, we conclude that the two non-equivalent
approximate eigenvalue-eigenvector pairs of $\mathcal F$
are $(\tilde E_\pm,c_\pm)$, 
where 
\bean
c_{\pm,lm} = \left\{
\begin{array}{ll}
	\tilde c_{\pm,1} + o(\epsilon^2) & \mbox{if $(l,m)=(\beta,-1)$,} \\
	\tilde c_{\pm,2} + o(\epsilon^2) & \mbox{if $(l,m)=(\alpha,0)$,} \\
	o(\epsilon) & \mbox{otherwise.}
\end{array}
\right. 
\eean
We neglect the perturbative corrections $\sim o(\epsilon), o(\epsilon^2)$ in
the eigenvectors $c_\pm$, and this implies that the approximate transition
probability is given by Eq. \eqref{eq:Pmrfundamental}.
Equation \eqref{eq:Pmrfundamental} predicts that 
complete Rabi oscillations are induced when $\Delta = 0$; 
solving Eq.~\eqref{eq:delta_fn}  for $\omega$ (recall that $\delta=\omega-\tilde B/\hbar$) provides the resonance frequency shown in Eq. \eqref{eq:mrfhres}.
The second term of Eq. \eqref{eq:mrfhres} 
corresponds to the Bloch-Siegert shift of the resonance frequency:
as the drive strength $\tilde B_\textrm{ac}$ is increased, the resonance frequency shifts
upwards. 
This feature is further discussed in Sec. \ref{sec:mrdiscussion}.
Finally, the Rabi frequency at the fundamental resonance, which is
given by $\hbar \Omega^{(1)}_\textrm{res} = 2\lambda$,
is expressed using Eq. \eqref{eq:lambda}
in Eq. \eqref{eq:mrfhrabi};
the result is the same as in the RWA. 



\subsubsection{Half-harmonic resonance}
\label{sec:mrhalfharmonic}

\begin{figure}[h]
\includegraphics[width=0.95\columnwidth]{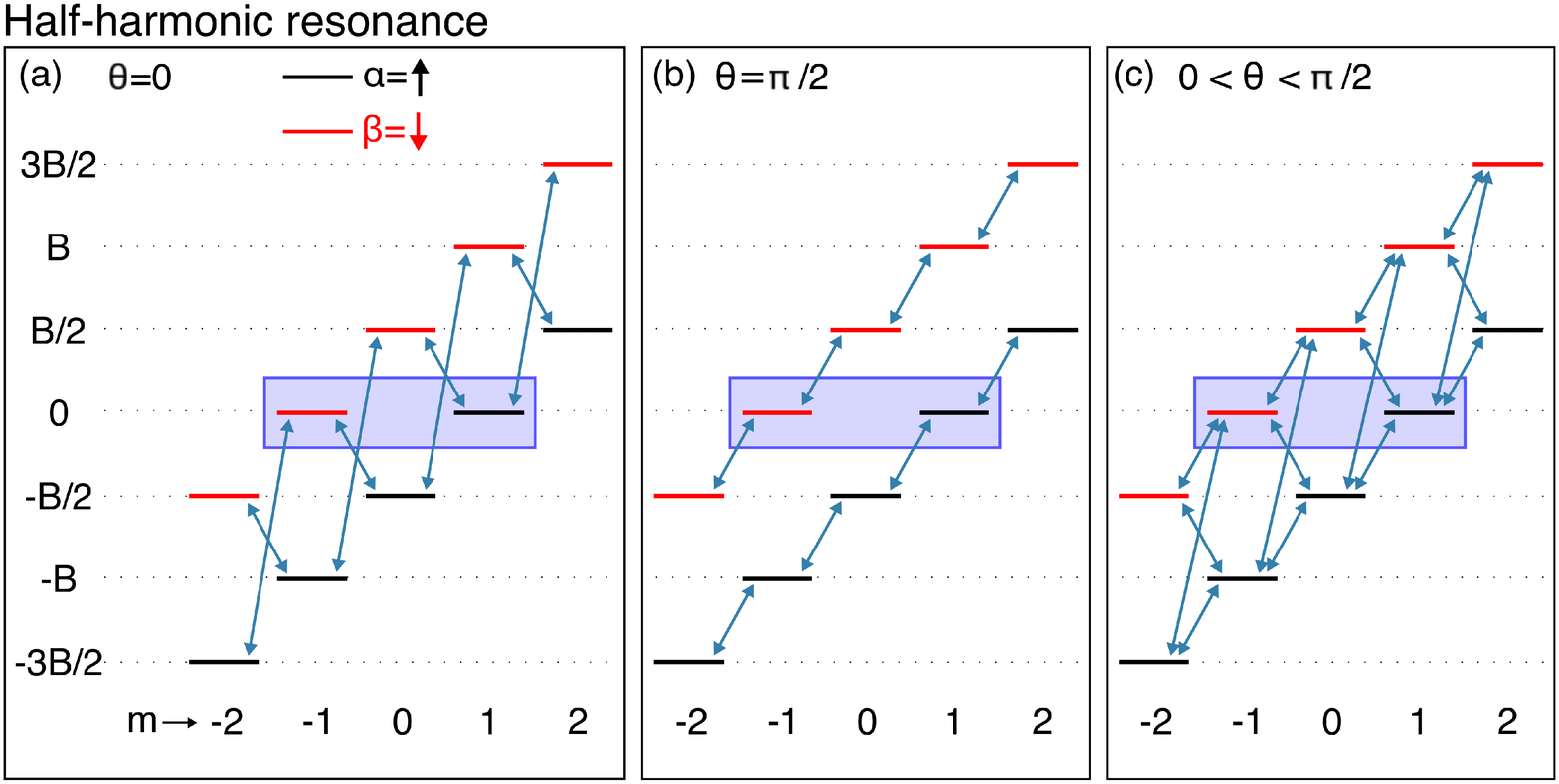}
\caption{
Magnetic resonance in a misaligned ac field: 
structure of the Floquet Hamiltonian at the half-harmonic resonance.
 Panels show cases when the ac field
 is perpendicular to the static field (a),
 is parallel to the static field (b),
 has finite perpendicular and parallel components (c).
 Horizontal lines (blue arrows) correspond to diagonal (off-diagonal) 
 matrix elements of the  Floquet Hamiltonian $\mathcal{F}$.
 The vertical position of each horizontal line corresponds to
 the value of the diagonal matrix element. 
}
\label{fig:mrhalfharmonic}
\end{figure}

Let us now consider the spin dynamics 
at half-harmonic resonance, when
$\hbar \omega \approx  \tilde B/2$. 
The level diagram visualising the Floquet Hamiltonian 
in the case $\hbar \omega = \tilde B/2$
is shown in Fig. \ref{fig:mrhalfharmonic}.
Again, we can identify degenerate pairs of Floquet levels, 
e.g., the pair $(\mathcal F_{\beta,-1,\beta,-1},\mathcal F_{\alpha,1,\alpha,1})$
highlighted with the blue box in Fig. \ref{fig:mrhalfharmonic}.

Note that in this case, there is no direct matrix element (blue arrow)
connecting these two Floquet levels. 
This implies that by repeating the  
first-order perturbation theory (equivalent to RWA)
done in Sec. \ref{sec:mrfundamentalRWA}, 
we would conclude that
two-photon Rabi oscillations do not happen.
However, this result is not correct; two-photon Rabi oscillations
can happen. 
To see that, we perform a 
second-order Schrieffer-Wolff transformation on $\mathcal F$,
as done in Sec. \ref{sec:mrfundamentalbeyondRWA}.
Furthermore we use the appropriate notation 
$\delta=\omega-\frac{\tilde B}{2\hbar}$. The obtained effective $2\times 2$ Floquet Hamiltonian $\tilde{\mathcal F}$ 
has the form given in Eq. \eqref{eq:gen2lev}, 
with 
\begin{subequations}
\bean
\label{eq:Delta}
\Delta &=&
	\hbar \delta -\frac{\tilde B_\textrm{ac}^2 \cos ^2 \theta}{6 \tilde B}, \\
\label{eq:lambdahalfharmonic}
\lambda &=& \frac{\tilde B_\textrm{ac}^2 \sin 2\theta}{8 \tilde B}.
\eean
\end{subequations}

Following the approach used in Sec. \ref{sec:mrfundamentalbeyondRWA},
and solving $\Delta=0$ we find that the half-harmonic resonance frequency 
$\omega^{(2)}_\textrm{res}$ is
given by Eq. \eqref{eq:mrhhres}.
Furthermore, using $ \hbar \Omega^{(2)}_\textrm{res} = 2 \lambda $ and
Eq. \eqref{eq:lambdahalfharmonic},
the Rabi frequency at the half-harmonic resonance
is obtained as shown in Eq. \eqref{eq:mrhhrabi}.


\subsection{Discussion}
\label{sec:mrdiscussion}

Let us now discuss the main features of the results
\eqref{eq:mrfhres},
\eqref{eq:mrfhrabi},
\eqref{eq:mrhhres}, and
\eqref{eq:mrhhrabi}.

Consider first the fundamental resonance frequency $\omega_\textrm{res}^{(1)}$
expressed in Eq. \eqref{eq:mrfhres}.
The second term in Eq. \eqref{eq:mrfhres} 
implies that $\omega_\textrm{res}^{(1)}$ has a positive drive-strength-dependent
correction $\propto \tilde B_\textrm{ac}^2/\tilde B$ with respect to the nominal Zeeman
splitting $\tilde B$. 
This correction is known as the BSS, which can be regarded as a special
case of the ac Stark shift\cite{Wei-BlochSiegertStark}.

Note that the parameter $\lambda$ and 
hence the Rabi frequency $\Omega^{(1)}_\textrm{res}$ 
sets the frequency broadening
of the fundamental  transition, 
as indicated by 
the prefactor $\lambda^2/(\lambda^2+\Delta^2)$
on the right hand side of Eq. \eqref{eq:Pmrfundamental}.
According to Eq. \eqref{eq:mrfhrabi}, this 
\emph{power broadening} of the fundamental resonance
is greater by a factor of $\tilde B/\tilde B_\textrm{ac}$ than the BSS.

Equation \eqref{eq:mrfhres} also shows that the BSS 
is finite in the limit of purely transversal drive ($\theta = 0$), 
and vanishes in the limit of purely longitudinal drive ($\theta = \pi/2$).
The respective Floquet level diagrams in 
Fig. \ref{fig:mrfundamental}a and b provide a
straightforward interpretation: 
the BSS can be regarded as a consequence of coupling-induced repulsion
between the Floquet levels. 
In Fig. \ref{fig:mrfundamental}a ($\theta = 0$), the Floquet level
$\mathcal F_{\beta, -1,\beta, -1}$ is connected by a  blue arrow
(off-diagonal matrix elements of $\mathcal F$) 
to the lower-lying Floquet level $\mathcal F_{\alpha,-2,\alpha,-2}$.
The consequence of this coupling in second-order perturbation theory
is level repulsion; i.e., the lower-lying Floquet level
pushes $\mathcal F_{\beta,-1,\beta,-1}$ upwards.
Similarly, $\mathcal F_{\alpha0,\alpha0}$ is pushed
downwards by its coupling to the  
higher-lying Floquet level
$\mathcal F_{\beta1,\beta1}$. 
These second-order level shifts appear in Eq. \eqref{eq:mrfundamentalDelta} as
the last term, and give rise to a finite BSS. 
In contrast, each of the highlighted Floquet levels in Fig. \ref{fig:mrfundamental}b
($\theta = \pi/2$) is connected to one higher-lying 
and one lower-lying Floquet level, and the
corresponding downward and upward level repulsions cancel each other,
giving rise to a vanishing BSS in this case.

Consider now the half-harmonic resonance. 
Equation \eqref{eq:mrhhrabi} provides the corresponding
Rabi frequency,
and it indicates the existence of 
Rabi oscillations unless $\theta = 0$ or $\theta = \pi/2$.
I.e., Rabi oscillations appear at  half-harmonic excitation 
only if  the transversal and longitudinal components of the 
driving field are both nonzero. 
The Floquet level diagrams shown in Fig. \ref{fig:mrhalfharmonic}
provide a visual interpretation of this feature:
Rabi oscillations arise if the blue arrows 
(off-diagonal matrix elements of $\mathcal F$)
draw at least one path between the two Floquet
levels highlighted by the purple box,
via virtual intermediate Floquet levels outside the box.
In the special cases $\theta = 0$ and $\theta = \pi/2$
depicted in Figs. \ref{fig:mrhalfharmonic}a 
and b, respectively,
no such paths exist.
However, there exist infinitely many such paths
for $0<\theta < \pi/2$ (Fig. \ref{fig:mrhalfharmonic}c), 
due to the coexistence of spin-conserving
and spin-flip off-diagonal matrix elements.
In particular, in our second-order Schrieffer-Wolff transformation
leading to 
the result \eqref{eq:mrhhrabi}, 
the two two-step paths via $\mathcal F_{\alpha,0,\alpha,0}$ 
and $\mathcal F_{\beta,0,\beta,0}$ are incorporated.

In the case of the half-harmonic resonance, 
the relation between the power broadening and the BSS is
qualitatively different from the case of the fundamental resonance. 
For the half-harmonic resonance, the power broadening is given by
Eq. \eqref{eq:mrhhrabi}, whereas the BSS is given by 
the second term of Eq. \eqref{eq:mrhhres}, i.e., 
the two quantities are of the same order, both being 
$\sim \tilde B_\textrm{ac}^2/\tilde B$.
Hence we expect that for the half-harmonic resonance, 
the BSS is relatively easily resolvable experimentally, 
at least if the dissipative frequency scales are smaller than the
power broadening.

Equation \eqref{eq:mrhhres} also shows that the BSS 
is finite in the limit of purely transversal excitation ($\theta = 0$), 
and vanishes in the limit of purely longitudinal excitation ($\theta = \pi/2$).
An interpretation completely analogous to the 
case of the fundamental resonance can be given 
based on the Floquet level diagrams in 
Fig. \ref{fig:mrhalfharmonic}a and b.

\section{Electrically driven spin resonance}
\label{sec:edsr}

\subsection{The model}

\label{sec:toymodel}

From now on, we describe EDSR
mediated by spin-orbit interaction in a 1D parabolic quantum dot. 
The setup is shown in Fig. \ref{fig:setup}.
The Hamiltonian 
\bean
\label{eq:edsrhamiltonian}
\mathcal H = \mathcal{H}_0 + \mathcal{H}_E + \mathcal{H}_B + \mathcal{H}_{\rm SO}
\eean
includes the harmonic-oscillator Hamiltonian ($\mathcal{H}_0$)
consisting of the kinetic energy of the electron and the parabolic confinement 
potential, the ac electric 
potential arising from the driving electric field ($\mathcal{H}_E$),
the static Zeeman effect caused by a homogeneous magnetic field ($\mathcal{H}_B$),
and the spin-orbit term ($\mathcal{H}_{\rm SO}$).
The explicit forms of these terms, respectively, are as follows:
\begin{subequations}
\begin{eqnarray}
\mathcal{H}_0 &=& \frac{p_z^2}{2m} + \frac 1 2 m \omega_0^2 z^2=\hbar\omega_0
\left(a^\dagger a^{\phantom{\dagger}}+\frac{1}{2}\right), 
\label{eq:H0}\\
\mathcal{H}_E &=& e z E_{\rm ac} \sin(\omega t)=\tilde  E_{\rm ac} \sin(\omega t)(a^\dagger+a^{\phantom{\dagger}}),
\\
\mathcal{H}_B &=& - \frac 1 2 g^* \mu_B B \sigma_z = -\frac 1 2 \tilde B \sigma_z
\\
\mathcal{H}_{\rm SO} &=& \alpha p_z \mathbf{n}_{\rm so} \cdot \boldsymbol{\sigma}
=
i \tilde \alpha (a^\dag - a) \mathbf{n}_{\rm so} \cdot \boldsymbol{\sigma}
\label{eq:HSO}
\end{eqnarray}\label{eq:model}
\end{subequations}
Here, $a$ and $a^\dag$ are the ladder operators of the
harmonic oscillator Hamiltonian, 
and $\mathbf{n}_{\rm so} = (0,\cos \theta, \sin \theta)$ is the 
direction of the effective magnetic field arising from spin-orbit coupling. 
Furthermore, we defined
\begin{eqnarray}
\tilde B &=& g^* \mu_B B  \\
\tilde \alpha &=&  
\alpha  \sqrt{\frac{m  \hbar \omega_0 }{2}}
\\
\tilde E_{\rm ac} &=& e E_{\rm ac} \sqrt{ \frac{\hbar}{2m\omega_0}}
\end{eqnarray}
These quantitites have the dimension of energy. 

Note that we use the same notation $\theta$ for two different quantities:
$\theta$ appears in Eq. \eqref{eq:mrfield} as the ac field misalignment angle in MR, 
and it also appears in this Section and in Fig. \ref{fig:setup}, as the angle characterising 
the direction of the spin-orbit term. 
We use the same notation for these quantities as they play very similar roles
in the spin dynamics. 

It is natural to represent the Hamiltonian terms \eqref{eq:model}
in the 
product basis of the orbital and spin degrees of freedom,
$\{ \ket{n_\sigma} \; | \;  n = 0,1,2,\dots; \, \sigma = \uparrow,\downarrow\}$,
where $n$ is the harmonic-oscillator orbital quantum number
and $\sigma$ is the spin quantum number with quantization
along $z$.

We will refer to the two lowest-energy eigenenstates of our
static Hamiltonian $\mathcal{H}_0 + \mathcal{H}_B + \mathcal{H}_{\rm SO}$ as the 
\emph{qubit basis states}.
The qubit basis state with the lower (higher) energy will be 
denoted by $\ket{G}$ ($\ket{E}$).

The electron is initialized in state $\ket{G}$ at $t=0$.
Our aim is to describe the time evolution of the state 
upon driving.
In particular, we are interested in the time-dependent
occupation probability $P_E(t)$ of state $\ket{E}$.
It is expected that at resonant driving $\hbar \omega \approx \tilde B$,
the dynamics resembles Rabi oscillations. 
Subharmonic (multi-photon or $N$-photon) 
resonances at $\hbar \omega \approx \tilde B /N$ (where $N = 1,2,\dots$) 
are also expected.
In this work we focus on the fundamental (single-photon, $N=1$) and
half-harmonic (two-photon, $N=2$)  resonances.

We aim at an analytical, perturbative description of spin transitions induced by 
the ac electric field. 
In particular, we calculate the resonance frequency and
the Rabi frequency at resonant driving. 
We consider the parameter range where the 
energy scale $\hbar \omega_0$
of the confinement potential dominates the other four energy scales,
the latter ones being assumed to be comparable in magnitude:
\begin{equation}
\label{eq:energyscales}
\hbar \omega \sim \tilde \alpha \sim \tilde E_{\rm ac} \sim \tilde B \ll \hbar \omega_0.
\end{equation}
This hierarchy of energy scales will allow for a 
perturbative description of the dynamics, with the small parameter
$\epsilon \sim \frac{\omega}{\omega_0} \sim \frac{\tilde \alpha}{\hbar \omega_0}
\sim \frac{\tilde E_\textrm{ac}}{\hbar \omega_0} \sim \frac{\tilde B}{\hbar \omega_0} \ll 1$.

\subsection{Effective qubit Hamiltonian}
\label{sec:qubitH}

In the EDSR problem defined in Sec. \ref{sec:toymodel},
the hierarchy of the energy scales is given by Eq. \eqref{eq:energyscales}.
Because of this hierarchy, an effective time-dependent two-level 
Hamiltonian [see Eq. \eqref{eq:edsreffectiveh} below] 
can be derived for the qubit dynamics,
using TDSW perturbation theory, which we outline in Appendix  \ref{sec:tdsw}.
This qubit Hamiltonian can then be used
to express the 
resonance frequencies 
$\omega_{\rm res}^{(1)}$ and $\omega_{\rm res}^{(2)}$,
and the corresponding Rabi frequencies at these resonances, 
$\Omega_{\rm res}^{\rm (1)}$ and
$\Omega_{\rm res}^{\rm (2)}$,
corresponding to the fundamental and half-harmonic resonances,
respectively [see Eqs. \eqref{eq:fund_res},
\eqref{eq:fund_res},
\eqref{eq:halfharmonicresonancefrequency},
and \eqref{eq:RabiAtHalfHarmonic} below].

We use the orbital-spin product basis 
$\{ \ket{n_\sigma} \; | \;  n =0,1,2,\dots; \, \sigma = \uparrow,\downarrow\}$,
as the starting point of TDSW,
and take the two-dimensional subspace
of $\ket{0_\uparrow}$ and $\ket{0_\downarrow}$ as
the relevant subspace in  TDSW.
We carry out a fifth-order TDSW (in the small parameter $\epsilon$), 
which is expected to describe
both the fundamental and the half-harmonic resonances.
The TDSW procedure 
yields the effective qubit Hamiltonian
\begin{equation}
\label{eq:edsreffectiveh}
\tilde{\mathcal H}_q \approx 
\tilde{\mathcal H}^{(0)}_q +
\tilde{\mathcal H}^{(1)}_q +
\tilde{\mathcal H}^{(2)}_q+
\tilde{\mathcal H}^{(3)}_q+
\tilde{\mathcal H}^{(4)}_q+
\tilde{\mathcal H}^{(5)}_q,
\end{equation}
where 
the six terms, representing terms from different orders in the perturbation, are listed
below in Eq. \eqref{eq:qubith05}.
Note that the terms
$\tilde{\mathcal{H}}_q^{(0)}$,
$\tilde{\mathcal{H}}_q^{(2)}$, and
$\tilde{\mathcal{H}}_q^{(4)}$ 
are proportional to the $2\times 2$ unit matrix $\sigma_0$, 
therefore they do not influence the dynamics, and hence
we disregard them in the forthcoming calculations; 
nevertheless we include them here for completeness:
\begin{subequations}
\label{eq:qubith05}
\begin{eqnarray}
\tilde{\mathcal H}^{(0)}_q  &=& \frac{\hbar \omega_0}{2} \sigma_0\;\\
\tilde{\mathcal H}^{(1)}_q &=& -\frac{\tilde B}{2} \sigma_3\;\\
\tilde{\mathcal H}^{(2)}_q&=&-\frac{\tilde\alpha^2+\tilde E_{\rm ac}^2\sin^2(\omega t)}{\hbar \omega_0}\sigma_0\;\\
\label{eq:qubitH3}
\tilde{\mathcal H}^{(3)}_q &=&
	-\frac{\tilde B\tilde E_{\rm ac}\tilde \alpha\cos\theta}{\hbar^2\omega^2_0}
		\sin(\omega t)\sigma_1 
	\\
	&-&\frac{\tilde\alpha\cos\theta}{\hbar^2\omega^2_0}
		(\tilde E_{\rm ac}\hbar\omega\cos(\omega t)
	+\tilde B\tilde\alpha\sin\theta)\sigma_2
		\nonumber
	\\
	&+&\frac{\tilde\alpha}{\hbar^2\omega^2_0}
		(\tilde B\tilde\alpha\cos^2\theta
		-\tilde E_{\rm ac}\hbar\omega\sin\theta\cos(\omega t))\sigma_3\;
		\nonumber
\\
 \tilde{\mathcal H}^{(4)}_q&=&
	-\frac{(\tilde B \tilde\alpha\cos\theta)^2}{\hbar^3 \omega^3_0}
		\sigma_0\;,\label{eq:Heff4}
\\
 \tilde{\mathcal H}^{(5)}_q&=& 
 	- \left(h^{(5)}_x \sigma_x + h^{(5)}_y \sigma_y 
		+ h^{(5)}_z \sigma_z \right),
 	\label{eq:Heff5}
\end{eqnarray}\label{eq:qubitH}
\end{subequations} 
In Eq. \eqref{eq:Heff5}, we used 
\begin{subequations}
\begin{eqnarray}
h^{(5)}_x &=&
 	\frac{\tilde B \tilde \alpha \tilde E_{\rm ac} \cos\theta}{\hbar^4 \omega^4_0}
	(2\tilde\alpha^2- \tilde B^2)\sin(\omega t)
\\
h^{(5)}_y &=&
	\frac{\tilde E_{\rm ac}\tilde \alpha\omega^3\cos\theta}{\hbar\omega^4_0}
		\cos(\omega t)
		\nonumber
		\\
		&+&
		\frac{\tilde B \tilde\alpha^2\sin2\theta}{2\hbar^4 \omega^4_0}
		(\tilde B^2-\tilde\alpha^2+\tilde E_{\rm ac}^2\sin^2(\omega t))	\\
h^{(5)}_z &=& 
	\frac{\tilde E_{\rm ac}\tilde \alpha\omega^3\sin\theta}
		{\hbar\omega^4_0}\cos(\omega t)
		\nonumber
		\\
		&+&\frac{\tilde B \tilde\alpha^2\cos^2\theta}{\hbar^4 \omega^4_0}
		(\tilde B^2-\tilde\alpha^2+\tilde E_{\rm ac}^2\sin^2(\omega t))
	\label{eq:Heff5defs}.
\end{eqnarray}
\end{subequations}
Note that the upper index in, e.g., $\tilde{\mathcal H}_q^{(3)}$ refers
to the order of perturbation theory in which the term appears.

Out of the six terms in 
Eq. \eqref{eq:edsreffectiveh}, 
$\tilde{\mathcal H}^{(0)}_q$ and  $\tilde{\mathcal H}^{(1)}_q$ are simply 
 the projected parts of $\mathcal{H}_0$ and $\mathcal{H}_1 \equiv
 \mathcal H_E + \mathcal H_B + \mathcal H_\textrm{SO}$, respectively. 
 $\tilde{\mathcal H}^{(2)}_q$ contains a static and a time-dependent
 second-order energy shift, 
due to the spin-orbit interaction and the ac electric field,
respectively. 
$\tilde{\mathcal H}^{(3)}_q$ has five terms. 
The first, second and fifth terms are spin- and time-dependent, 
hence these all contribute to the qubit dynamics. 
The third and fourth terms are static; they 
describe the spin-orbit-induced g-tensor renormalization.
The fourth-order term $\tilde{\mathcal{H}}_{q}^{(4)}$ 
of the qubit Hamiltonian, being diagonal, does not influence spin dynamics.
The static parts of the fifth-order term $\tilde{\mathcal{H}}_{q}^{(5)}$ 
describe higher-order g-tensor renormalisation (those
proportional to $\tilde \alpha^4 \tilde B$), 
or nonlinear Zeeman splitting (those proportional to $\tilde \alpha^2 \tilde B^3$).

Already at this point, there are reasons to expect that in 
this EDSR model, 
a half-harmonic resonance occurs, and that 
the half-harmonic resonance frequency is driving-strength dependent:
(i) The third-order effective Hamiltonian $\tilde{\mathcal{H}}_q^{(3)}$ 
incorporates both longitudinal and transverse ac components, 
in analogy with the case of the misaligned-field MR discussed 
in Sec. \ref{sec:esr}.
(ii) The fifth-order effective Hamiltonian $\tilde{\mathcal H}_q^{(5)}$ 
incorporates terms proportional to 
$\tilde{E}^2_\textrm{ac} \sin^2\omega t
= \frac 1 2 \tilde{E}^2_{\textrm ac} (1-\cos 2 \omega t)$.
The longitudinal static part $\propto \tilde E^2_\textrm{ac} \sigma_z$
can be interpreted as a drive-strength-dependent effective 
g-tensor renormalisation, which contributes to the BSS, 
whereas the dynamical part 
$\propto \tilde E^2_\textrm{ac} \cos 2 \omega t \ \sigma_y$
is expected to drive Rabi oscillations at half-harmonic
excitation, i.e., when $2\hbar \omega \approx \tilde B$.

We note that the effective Hamiltonian in Eq. \eqref{eq:qubitH}
fulfills the expectation that no spin transition
occurs if the external B-field and the spin-orbit field are
aligned, ie, when $\theta=\pi/2$.

\subsection{Floquet perturbation theory for EDSR}

We apply  Floquet perturbation theory, outlined in 
Sec. \ref{sec:mrfloquetperturbation},
to describe the fundamental and half-harmonic
resonances. 
In particular, we derive the parameter dependence of the 
corresponding resonance frequencies $\omega_\textrm{res}^{(1)}$ and
$\omega_\textrm{res}^{(2)}$,
as well as the Rabi frequencies 
$\Omega_\textrm{res}^{(1)}$ and 
$\Omega_\textrm{res}^{(2)}$,
at these two resonances,
up to terms of the order of $\sim \tilde B \epsilon^4$. 
There are two significant differences in the derivation of the EDSR results
with respect to that of the MR results; we outline these
differences in the following. 

(1) The MR Hamiltonian \eqref{eq:ham_esr}
 has a driving term that is proportional to $\sin \omega t$. 
In contrast, the effective qubit Hamiltonian \eqref{eq:edsreffectiveh}
we obtained for EDSR 
has $\cos \omega t$ terms as well as second-harmonic terms
proportional to $\cos 2 \omega t$. 
In practice, the latter fact implies that the Floquet matrix 
will contain off-diagonal matrix elements that connect
Floquet levels with next-nearest-neighbor Floquet quantum numbers. 

(2) In the EDSR case, we repeat the same second-order time-indepedent
Schrieffer-Wolff transformation on the Floquet Hamiltonian $\mathcal F$
that we applied in 
Secs. \ref{sec:mrfundamentalbeyondRWA} and \ref{sec:mrhalfharmonic}.
The Floquet Hamiltonian itself contains terms
of the order of $~\tilde B$, $~\tilde B \epsilon^2$ and $~ \tilde B \epsilon^4$, since
it is constructed from the effective qubit Hamiltonian that is itself the result
of a finite-order perturbative calculation.
When we separate the Floquet Hamiltonian to diagonal ($\mathcal F_0$)
and off-diagonal ($\mathcal F_1$) components, and
apply  time-independent Schrieffer-Wolff transformation
up to second order in $\mathcal F_1$, 
the resulting $2\times 2$ effective Floquet Hamiltonian 
will involve higher-order terms, up to $~\tilde B \epsilon^8$. 
As our original Hamiltonian was accurate only up to 
the $\sim \tilde B \epsilon^4$ terms, we  drop the terms that are of higher order than $\sim \tilde B \epsilon^4$ from the effective Floquet
Hamiltonian.

\subsection{Analytical vs. numerical solution}\label{sec:results}

The results we obtain from Floquet perturbation theory are shown in 
Sec. \ref{sec:summaryofresults} as
Eqs. \eqref{eq:fund_res}--\eqref{eq:RabiAtHalfHarmonic}.
In the rest of this subsection we discuss these results and compare them to 
numerical results.


The terms 
describing the fundamental resonance frequency
in Eq. \eqref{eq:fund_res}
are interpreted as 
 nominal Zeeman splitting, 
g-tensor renormalisation,
nonlinear Zeeman effect, 
and BSS, respectively.
We call the last term a BSS as it is
a power-dependent correction to the resonance frequency, 
that is second order in the drive amplitude, hence
analogous to the BSS in MR.
Remarkably, the BSS in Eq. \eqref{eq:fund_res}
is a \emph{negative} correction, whereas the BSS in MR is always positive.
The last term of 
the half-harmonic resonance frequency 
[Eq. \eqref{eq:halfharmonicresonancefrequency}]
is also interpreted as a BSS. 
Further similarities with the MR case: 
(i) For the fundamental resonance, the BSS is smaller ($\sim \tilde B\epsilon^4$)
than the power broadening, the latter being given by 
$\hbar \Omega_\textrm{res}^{(1)} \sim \tilde B \epsilon^2$.
(ii) For the half-harmonic resonance, the BSS, 
being $\sim \tilde B \epsilon^4$, is comparable 
to the power broadening, the latter being given by 
$\hbar \Omega_\textrm{res}^{(2)} \sim \tilde B \epsilon^4$. 
(iii) The BSS for both the fundamental and the half-harmonic resonance
is proportional to $\cos^2 \theta$, i.e., it vanishes in the limit
of purely longitudinal excitation, and finite for purely transversal excitation. 
These features can be explained by the argument provided in 
Sec. \ref{sec:mrdiscussion} for the case of MR, applied
to the effective qubit Hamiltonian \eqref{eq:edsreffectiveh}.

Regarding the results \eqref{eq:fund_res} and
\eqref{eq:halfharmonicresonancefrequency} for the resonance frequencies,
we note that their ratio is exactly two in the limit of vanishing driving power, 
i.e., $\lim_{\tilde E_\textrm{ac} \to 0} \left( 
\frac{\omega_\textrm{res}^{(1)}}{\omega_\textrm{res}^{(2)}}
\right) = 2$.

We have checked that the result \eqref{eq:fund_res} for 
the fundamental Rabi frequency $\Omega^{(1)}_\textrm{res}$
matches the corresponding result of 
Ref. \onlinecite{Golovach-edsr}; see Appendix \ref{app:gbl} for details.

\begin{figure}
\includegraphics[width=1\columnwidth]{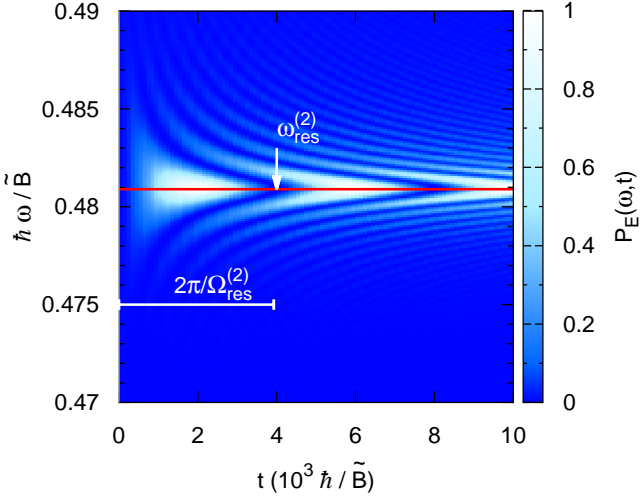}
\caption{(color online) Electrically driven Rabi spin dynamics
at half-harmonic resonance. 
The excited-state occupation probability
$P_E$ is shown as a function of the
drive frequency $\omega$ and time $t$;
the numerical data reveals the chevron pattern 
characteristic of magnetic resonance. 
Parameters: $\theta = \pi/4$,
$\tilde \alpha/\tilde B = \tilde E_{\rm ac}/\tilde B =1$,
$\hbar \omega_0/ \tilde B = 5$. 
The analytical results for the half-harmonic resonance 
frequency $\omega^{(2)}_\textrm{res}$ and the Rabi frequency $\Omega^{(2)}_\textrm{res}$ are also displayed.
For the above parameter values, the latter one is related to the time period
of the oscillation via $2\pi/\Omega^{(2)}_\textrm{res} =
2\pi \times 625\, \hbar / \tilde B$ .}
\label{fig:chevron}
\end{figure}

The analytical results are tested against numerically exact solutions
of the time-dependent Schr\"odinger equation defined by the Hamiltonian
$\mathcal H$ in Eq. \eqref{eq:edsrhamiltonian}. 
The numerical results were obtained using the truncated Hilbert space
spanned by the 8 lowest-energy eigenstates of $\mathcal H_0 + \mathcal H_B$, 
corresponding to the 4 lowest-lying levels of the harmonic oscillator.
We have checked that there was no visible change in the numerical results
upon extending the Hilbert space with further, higher-lying orbitals.

In Fig. \ref{fig:chevron}, we plot the numerically
computed time evolution of the
occupation probability of the excited state $\ket{E}$, for a finite range
of the driving frequency in the vicinity of the `nominal'
half-harmonic resonance frequency $\hbar \omega / \tilde B = 0.5$
(see caption for parameter values).
The analytical result \eqref{eq:halfharmonicresonancefrequency} predicts 
complete Rabi oscillations at $ \omega = \omega^{(2)}_\textrm{res} =
0.4809 \tilde B/\hbar$,
and the Rabi frequency at this resonance is predicted by Eq. \eqref{eq:RabiAtHalfHarmonic} to be 
$\Omega^{(2)}_\textrm{res} \approx \frac{1}{625}\frac{ \tilde B}{\hbar}$. 
These predictions are  in line with the numerical data shown in Fig. \ref{fig:chevron}. 
For a finite detuning from the resonance frequency,
the Rabi oscillations become faster and reduced
(i.e., they do not reach $P_E =1$), leading to the 
characteristic chevron pattern\cite{Kawakami-edsr,Veldhorst} 
known from MR.
The results of Fig. \ref{fig:chevron} therefore reveal simple Rabi
dynamics at the half-harmonic resonance.

\begin{figure}
\includegraphics[width=1\columnwidth]{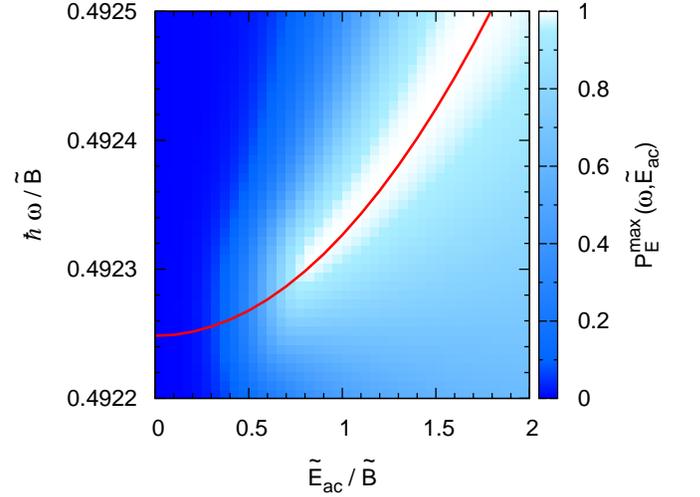}
\caption{(color online) Bloch-Siegert shift and power broadening 
of the half-harmonic resonance.  
The maximal excited-state occupation probability
$P^{\rm max}_E$ is shown as a function of 
the amplitude $\tilde E_{\rm ac}$ of the driving ac electric field
 and drive frequency $\omega$.
Parameters: $\theta = \pi/4$
$\tilde \alpha / \tilde B = 1$, and
$\hbar \omega_0 / \tilde B = 8$. The red line indicates the analytical result for the resonance frequency as the function of electric field based on Eq.~\eqref{eq:halfharmonicresonancefrequency}.}
\label{fig:PE_E_omega}
\end{figure}

The density plot of Fig. \ref{fig:PE_E_omega} 
is a visual demonstration of the BSS, i.e., of that 
the resonance frequency increases with increasing 
drive strength. 
The figure shows the maximum $P_E^{\rm max}$ 
of the excited-state
probability $P_E(t)$ within a time span exceeding
the Rabi period at the half-harmonic resonance,
as a function of the amplitude $E_{\rm ac}$ of the 
ac electric field and the drive frequency $\omega$. 
(See caption for parameters.)
Therefore, vertical cuts of the density plot correspond to 
resonance curves. 
The solid line represents the analytical result \eqref{eq:halfharmonicresonancefrequency}
for the 
half-harmonic resonance frequency.
The agreement between the 
analytical curve and the $P_E^{\rm max} \approx 1$
ridge of the numerical simulation reassures  the validity 
and correspondence of the two approaches. 
Importantly, in Fig. \ref{fig:PE_E_omega},
the BSS is comparable in magnitude to 
the power broadening, which makes the BSS
relatively easily resolvable in experiments realizing the
model we use.

\begin{figure}
\includegraphics[width=1\columnwidth]{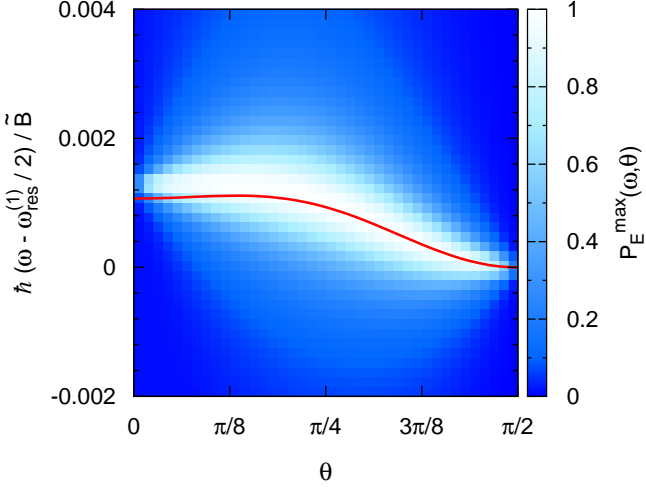}
\caption{(color online) 
Anisotropy of the half-harmonic resonance. 
The maximal excited-state occupation probability $P_E^{\rm max}$ 
is shown as a function of the angle $\theta$ characterizing 
the spin-orbit interaction and the drive frequency $\omega$
measured from $\omega_{\rm res}^{(1)}/2$, see Eq. \eqref{eq:fund_res}. 
The solid line corresponds to the analytically obtained
half-harmonic resonance frequency, i.e., it shows
$\omega^{(2)}_{\rm res}-\frac{\omega^{(1)}_{\rm res}}{2}$, see 
Eq.~\eqref{eq:halfharmonicresonancefrequency}.
Parameters: 
$\tilde E / \tilde B = 1$, $\tilde \alpha / \tilde B = 1$, and
$\hbar \omega_0 / \tilde B = 5$.
}
\label{fig:P_omega_theta}
\end{figure}

A further question is how the BSS depends on the 
angle $\theta$ characterizing the direction of the 
spin-orbit interaction.
This dependence is exemplified by Fig. \ref{fig:P_omega_theta},
which shows $P^{\rm max}_E$ as a function
of $\theta$ and the drive frequency. 
The latter is measured from half of the
calculated fundamental
resonance frequency $\omega_{\rm res}^{(1)}$,
see Eq. \eqref{eq:fund_res}.
The solid line, showing good agreement with
the centre of the bright $P_E^{\rm max} \approx 1$
region of the underlying density plot, shows the analytical result
for the half-harmonic resonance frequency $\omega_{\rm res}^{(2)}$
[Eq. \eqref{eq:halfharmonicresonancefrequency}].

\section{Discussion and conclusions}
\label{sec:conclusions}

(1) We provide a numerical example to estimate orders of magnitudes
of the EDSR resonance shifts and Rabi frequencies. 
Let us take $B = 1.7$ T with the electronic $g$ factor 2, yielding
$\tilde B \approx 0.1$ meV. 
We set $\tilde \alpha = 0.1$ meV and $\tilde E_\textrm{ac} = 0.1$ meV,
and the orbital level spacing is chosen to be $\hbar \omega_0 = 1$ meV. 
Then the order of magnitude of the Rabi frequency 
at the fundamental resonance
becomes $\Omega_\textrm{res}^{(1)}\sim \tilde B \epsilon^2 / \hbar \approx
1.5 \times 10^9 \frac{1}{\textrm{s}}$ 
corresponding to a spin-flip time of $\approx 4.3$ ns.
For the half-harmonic resonance, 
$\Omega_\textrm{res}^{(2)} \sim \tilde B \epsilon^4/ \hbar
\approx 1.5 \times 10^7 \frac{1}{\textrm{s}}$, 
corresponding to a spin-flip time of $\approx 430$ ns. 
For both resonances, the BSS is comparable to the value 
of $\Omega_\textrm{res}^{(2)}$ estimated above. 

(2) The results presented in this work describe a perturbative regime
where spin-orbit interaction is assumed to be `weak', in the sense
that the spin-orbit energy scale in the QD is dominated by the
QD level spacing, $\tilde \alpha \ll \hbar \omega_0$.
In nanowire QD host materials such as InAs\cite{Schroer-gfactor} 
and InSb\cite{NadjPerge}, spin-orbit interaction 
is known to be `strong' in the sense that it creates a strong g-factor renormalisation,
already in the bulk materials. 
A question arising from these facts is: are typical 
InAs and InSb nanowire QDs within the range of validity 
of our perturbative theory?
One way to answer this question is via a comparison of the 
dependence of the fundamental EDSR resonance frequency 
obtained from the perturbative theory and from experiments. 
The experiments\cite{Schroer-gfactor,NadjPerge} have found that the fundamental 
resonance frequency shows a similar angular dependence 
as the perturbative result Eq. \eqref{eq:fund_res}, i.e., 
for a magnetic field with a fixed magnitude, 
the resonance frequency is maximal if the magnetic field is aligned along a certain
direction and minimal if it is aligned perpendicular to that direction. 
To be specific, we take the data given in the first row of Table I. of 
Ref. \onlinecite{Schroer-gfactor}, which indicates that the ratio of the minimal and maximal
resonance frequencies in the considered case were $\approx 0.84$.
Using the first two terms in 
Eq. \eqref{eq:fund_res}, 
we can identify that ratio with $1-2\tilde \alpha^2/\hbar^2\omega_0^2$, 
yielding $\tilde \alpha / \hbar \omega_0 \approx 0.28$ 
for this particular InAs device. 
A similar analysis of the experimental data in Fig. 3c of Ref. \onlinecite{NadjPerge}
results in an estimate $\tilde \alpha / \hbar \omega_0 \approx 0.37$ 
for the measured InSb device. 
These estimates suggest that the InAs and InSb QDs are on the
border between `weak' and `strong' spin-orbit interaction.

(3)
To our knowledge, three experiments have reported subharmonic EDSR resonances
in semiconductor nanowire QDs, where our model based on the Rashba-type
spin-orbit interaction could be appropriate to describe the spin dynamics.
The strong subharmonic resonances reported in Stehlik et al.\cite{Stehlik-harmonic} 
are described by a theory developed for strongly driven double quantum dots
\cite{DanonRudner}. 
Faint half-harmonic resonances are  visible in the data of Refs. 
\onlinecite{Schroer-gfactor} (see Fig. 2b therein)
and 
\onlinecite{NadjPerge} (see Fig. 2b therein).
A quantitative 
experimental analysis exploring the parameter dependencies
of the corresponding resonance and Rabi frequencies
would allow for a comparison with our predictions.

(4) One of our conclusions was that the BSS of the fundamental EDSR resonance
frequency
has an anomalous, negative sign, see Eq. \eqref{eq:fund_res}.
Here, we provide a simple physical picture explaining 
this result, using the unitary transformation 
applied in Ref. \onlinecite{Levitov}. 
For simplicity, we focus on the case when the spin-orbit field is perpendicular
to the magnetic field, i.e., $\theta = 0$.  
Then, the unitary transformation  $S$ of Ref. \onlinecite{Levitov} 
(not to be confused with the generator of the Schrieffer-Wolff
transformation in Appendix \ref{sec:tdsw}) applied on
our static Hamiltonian $\mathcal H_0 + \mathcal H_B +\mathcal H_\textrm{SO}$
eliminates the spin-orbit term and transforms the homogeneous magnetic field
$H_B$ to an inhomogeneous, spiral-like magnetic field, 
$H'_B \equiv S H_B S^\dag 
\propto \tilde B \left[ \sigma_z \cos (z/\xi) - \sigma_x \sin(z/\xi) \right]$, 
where $\xi \propto 1/ \tilde \alpha$ is the spin-orbit length
[see Eq. (2) of Ref. \onlinecite{Levitov}].
The driving electric field, incorporated in our model as $\mathcal H_E$,
 induces a spatial oscillation $z(t) = -A \sin \omega t$ of the electron's
centre of mass with an amplitude $A \propto \tilde E_\textrm{ac}$.
Inserting this time-dependent $z(t)$ to the above expression for
$H'_B$, and expanding the terms up to second order in $A/\xi$, 
we find
$H'_B(t) \propto \tilde B \left[ \sigma_z \left(1-\frac{A^2}{\xi^2} \sin^2\omega t\right)
+ \sigma_x \frac{A}{\xi} \sin \omega t \right]$.
That is, the time-averaged z component of the time-dependent 
magnetic field in $H'_B(t)$ 
acquires a correction  
proportional to $ - \tilde B \frac{A^2}{\xi^2}
\propto - \tilde B \tilde E_\textrm{ac}^2 \tilde \alpha^2$. 
Notice that this correction is negative and has the same parameter dependence as
the BSS in, the last term of, Eq. \eqref{eq:fund_res}.
 
(5) To our knowledge, BSS has not yet been experimentally or theoretically analyzed in the context of EDSR. However, we wish to point out that certain numerical results in Ref.~\onlinecite{Khomitsky}, related to EDSR in a double quantum dot, are reminiscent of the BSS. Figures 4a and 4b in Ref.~\onlinecite{Khomitsky} show spin Rabi oscillations for different drive strengths. Therein, the drive strength is characterised by the dimensionless quantity $f$. In Figs. 4a and 4b of Ref.~\onlinecite{Khomitsky} it is shown that the complete Rabi oscillations at the fundamental resonance become incomplete upon increasing the drive strength from $f=0.02$ to $f=0.15$, while the driving frequency is maintained. This phenomenology is reminiscent of the effect of BSS: when the drive strength is increased, BSS provides a shift of the resonance frequency, hence a fixed drive frequency becomes off-resonant, and the Rabi-oscillation amplitude decreases. It is therefore tempting to interpret these results as consequences of BSS. However, the phenomenology of BSS would imply that (i) upon further increase of the drive strength, e.g., at $f=0.35$, the amplitude of the Rabi oscillation further decreases, and (ii) the Rabi oscillation speeds up gradually as $f$ is increased from $0.02$ to $0.15$ and to $0.35$. The results shown in Figs. 4a and 4b of Ref.~\onlinecite{Khomitsky} disagree with these expectations, hence we conclude that the BSS phenomenology is insufficient to describe the numerical results of Ref.~\onlinecite{Khomitsky}. Importantly, Ref.~\onlinecite{Khomitsky} considers parameter settings where the undriven system consists of four, approximately equidistant levels (see their Fig. 2a), which is a key difference with respect to the effectively two-level setup considered in our present work, and can be responsible for the phenomenology deviating from that of the BSS. 

(6) Even though EDSR experiments can be performed on single QDs\cite{Nowack-esr,Kawakami-edsr},
many current experiments use the Pauli blockade setup 
for initialization and readout\cite{Koppens-esr}.
The latter setup consists of a double QD which is occupied by 
two electrons (in the simplest case) during EDSR.
The inherent anharmonicity of the double QD confinement potential,
as well as the presence of the Coulomb interaction between the two (or more)
electrons, can provide alternative nonlinear EDSR mechanisms\cite{Rashba-subharmonic-edsr,Tokura,Khomitsky,Laird-sst,DanonRudner,Stehlik-harmonic}, 
which compete with
those presented in our work focusing on harmonic confinement and
single-electron dynamics. 
For example, an apparently well understood\cite{DanonRudner} 
case when the two-electron and double-QD features dominate the subharmonic 
EDSR resonances is the experiment of Ref. \onlinecite{Stehlik-harmonic}.

In conclusion, we 
have studied the characteristics of EDSR in a 1D QD model with parabolic 
confinement, homogeneous Rashba spin-orbit interaction and homogeneous
driving electric field. 
We demonstrated the existence of subharmonic (multi-photon)
resonances in this model, 
and analysed the half-harmonic (two-photon) resonance
in detail.
We have analytically described the parameter dependence of the
fundamental resonance frequency and the half-harmonic resonance 
frequency, and demonstrated that these resonance frequencies 
increase with increasing drive strength.
This effect is analogous to the BSS in MR. 

Our results describe a perturbative regime, where
the orbital level spacing
of the QD  dominates the energy scales of the external magnetic field,
spin-orbit interaction, and electrical drive. 
Therefore our results have direct 
experimental relevance for QDs with weak spin-orbit interaction.
They can also serve as benchmarks for numerical studies 
departing from the perturbative regime.
The model used here contains only minimal ingredients 
necessary to describe EDSR, suggesting that the subharmonic 
resonances and the BSS discussed here are generic features
of electrically driven spin dynamics.

\emph{Note added:} Upon completing this work, we became aware of a related
experiment\cite{Scarlino} revealing half-harmonic EDSR in a single QD, 
mediated by an inhomogeneous magnetic field.

\acknowledgments

We thank P. Scarlino and G. Sz\'echenyi for useful discussions. 
We acknowledge funding from the EU Marie Curie Career Integration Grant 
CIG-293834 (CarbonQubits), the OTKA Grants PD 100373 and 106047, 
and the EU ERC 
Starting Grant CooPairEnt 258789. 
GB acknowledges funding from the Deutsche Forschungsgemeinschaft (DFG)
within SFB767 and from the EU Marie Curie ITN S3NANO.
AP is supported by the J\'anos Bolyai Scholarship of the Hungarian Academy of Sciences.

\appendix

\section{Time-dependent Schrieffer-Wolff perturbation theory}
\label{sec:tdsw}

Here we introduce the time-dependent Schrieffer-Wolff
perturbation theory (TDSW), the method we use to derive the
effective $2\times 2$  time-dependent Hamiltonian 
\eqref{eq:edsreffectiveh} governing the
dynamics of the qubit. 
\begin{figure}
\includegraphics[width=0.9\columnwidth]{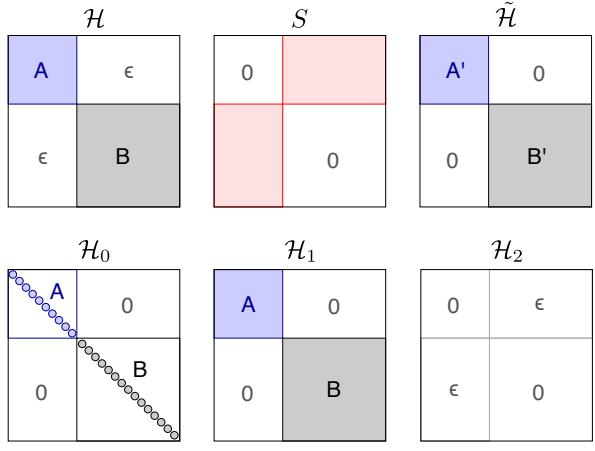}
\caption{Schematic form of the matrices we encounter during the SW perturbation theory.
The label $\epsilon$ refers to blocks
with matrix elements that are much smaller than the energy difference
between the relevant and irrelevant subspaces. 
In the EDSR problem, the energy scale of the those blocks
is $\tilde B  \sim \epsilon \hbar \omega_0$.}
\label{fig:SW}
\end{figure}

Let us first recall the basic idea of standard time-independent
Schrieffer-Wolff (SW) perturbation theory.~\cite{Schrieffer1966,Winkler}
We consider a Hamiltonian $\mathcal{H}=\mathcal{H}_0+\mathcal{H}'$ where $\mathcal{H}_0$ is diagonal and $\mathcal{H}'$ is the perturbation. Furthermore, 
 the basis states of $\mathcal{H}$ can be divided into relevant ${\sf A}$ and irrelevant ${\sf B}$ subspaces that have well separated energy scales. ${\sf A}$ and ${\sf B}$ are weakly interacting,  i.e. the matrix elements connecting them are small compared to the energy separation of the two subspaces. 
Ideally, we can introduce a unitary transformation $e^{-S}$ that brings $\mathcal{H}$ into a block diagonal form $\mathcal{\tilde H}=e^{-S}\mathcal{H}e^{S}$ where the relevant and irrelevant subspaces are separated as illustrated in Fig.~\ref{fig:SW}. 
However, in most of the cases we don't know the explicit form of the transformation $e^{-S}$ so we have to construct it bit-by-bit until the elements connecting the two subsets vanish up to the desired order of perturbation. This is usually done by expanding $e^{-S}$ in a series and constructing the terms of different orders successively. 

A great advantage of SW with respect to conventional perturbation theory is that here we don't need to distinguish between the degenerate and non-degenerate cases. 

Now we introduce the 
\emph{time-dependent} SW perturbation theory as a natural extension of the time-independent case. 
Similar approaches have been applied for
particular problems in Refs.~\onlinecite{Schwabl,Kaminski2000,Goldin2000};
here, we provide a general description of the method,
which we utilised in the main text for deriving the effective qubit Hamiltonian 
\eqref{eq:edsreffectiveh} of the EDSR problem.

%

Consider the time-dependent Hamiltonian 
$\mathcal H(t) = \mathcal H_0 + \mathcal H'(t)$,
where the perturbation is divided into a block-diagonal
and block-off-diagonal  part $\mathcal H'(t)=\mathcal H_1+\mathcal H_2$ as shown in Fig~\ref{fig:SW}.

In our problem (see Sec. \ref{sec:toymodel}), 
$\mathcal H_1 =  \mathcal{H}_B$ and
$\mathcal H_2(t) =  \mathcal{H}_E(t) + \mathcal H_{\rm SO}$.
Note that there $\mathcal H_1$
happens to be a time-independent perturbation,
but the treatment outlined here is readily applicable to a time-dependent
block-diagonal  perturbation as well. 

Similarly to the SW we successively build the unitary transformation $U(t) = e^{-S(t)}$ that separates the subspaces ${\sf A}$ and ${\sf B}$, but here the matrix $S(t)$ is now time-dependent. 
Note that any unitary transformation can be written in this form, and
the matrix $S(t)$ should be anti-Hermitian to ensure the unitary
character of $U(t)$.
The matrix $S(t)$ is chosen to be block-off-diagonal (see Fig~\ref{fig:SW}).
Note also that because of the weakness of the inter-subspace
coupling, the unitary tranformation $U(t)$ is close 
to unity, and hence $S(t)$ is small and can be expressed
as a power series with respect the perturbing terms. 

The transformation of
time-dependent Schr\"odinger equation 
$-i \hbar \frac{\partial}{\partial t} \psi(t) + \mathcal H(t) \psi(t) = 0$ with 
the above $U(t)$ is canonical, i.e., it preserves the form 
of the time evolution equation. 
The transformed wave function and Hamiltonian 
read as: 
\begin{equation}
\tilde \psi(t) = e^{-S(t)} \psi(t),
\end{equation}
\begin{eqnarray}
\mathcal{\tilde H}(t)=e^{-S(t)}\mathcal{H}(t) e^{S(t)}+i \hbar \frac{\partial e^{-S(t)}}{\partial t} e^{S(t)}.
\label{eq:tdsw}
\end{eqnarray}
From now on, we might suppress the time argument and
denote  time derivatives such as $\frac{\partial}{\partial t} \psi$
as $\dot \psi$. 

Starting from Eq. \eqref{eq:tdsw}, 
we utilize the power series of the exponential function. The second term in Eq.~ \eqref{eq:tdsw} is the heart of the \emph{time--dependent} SW transformation; in the time--independent case this term vanishes as $S$ is time-independent.
The expansion of the first term in Eq.~\eqref{eq:tdsw} is known from SW formalism, therefore we do not discuss it here.
The explicit form of the second term, after expanding the exponential function, has the following form:
\begin{widetext}
\begin{eqnarray}
\frac{\partial e^{-S}}{\partial t} e^{S}&\!=\!&
\left[\frac{\partial}{\partial t} \left(\!-\!S\!+\!\frac{1}{2!}S^2\!-\!\frac{1}{3!}S^3\!+\dots\!\right)\right]
(I\!+\!S\!+\!\frac{1}{2!}S^2\!+\!\frac{1}{3!}S^3\!+\dots\!)\!=\!(\!-\!\dot{S}\!+\!\frac{1}{2!}\dot{S}S\!+\!\frac{1}{2!}S\dot{S}\!-\!\frac{1}{3!}\dot{S}S^2\!-\!\frac{1}{3!}S\dot{S}S\!-\!\frac{1}{3!}S^2\dot{S}\!+\!\dots)\times\nonumber\\
&&\times(I\!+\!S\!+\!\frac{1}{2!}S^2\!+\!\frac{1}{3!}S^3\!+\!\dots)\!=\!(\!-\!\dot{S}\!+\!\frac{1}{2!}S\dot{S}\!-\!\frac{1}{2!}\dot{S}S\!-\!\frac{1}{3!}\dot{S}S^2\!+\!\frac{1}{3}S\dot{S}S\!-\!\frac{1}{3!}S^2\dot{S}\!+\dots\!)\nonumber\\
&=&-[\dot{S},S]^{(0)}-\frac{1}{2!}[\dot{S},S]^{(1)}-\frac{1}{3!}[\dot{S},S]^{(2)}\dots=-\sum_{j=0}^{\infty}\frac{1}{(j+1)!}[\dot{S},S]^{(j)}\;.
\end{eqnarray}
\end{widetext}
The transformed Hamiltonian then equals to
\begin{eqnarray}
\label{eq:transformed}
\mathcal{\tilde H}=\sum_{j=0}^{\infty}\frac{1}{j!}\left[\mathcal{H},S\right]^{(j)}-i \hbar\sum_{j=0}^{\infty}\frac{1}{(j+1)!}[\dot{S},S]^{(j)}\;.
\end{eqnarray}
with 
$\left[\mathcal{H},S\right]^{(n+1)} =\left[\left[\mathcal{H},S\right]^{(n)},S\right]$ and  $\left[\mathcal{H},S\right]^{(0)}=\mathcal{H}$.
Note that the second term in \eqref{eq:transformed}
is new with respect to time-independent SW, and it is a consequence of the time dependence of the Hamiltonian and therefore that of the matrix $S$. Considering a time-independent Hamiltonian the second term vanishes and we are left with the well-known SW transformation.

We now exploit the block-off-diagonal property of $S$ in order to 
separate the block-off-diagonal and block-diagonal parts of the transformed
Hamiltonian:
\begin{eqnarray}
\mathcal{\tilde H}_{\text{off-diag}} &=&
\sum_{j=0}^{\infty}\frac{1}{(2j +1)!}\left[\mathcal{H}_0+\mathcal{H}_1,S\right]^{(2j+1)} 
\nonumber
\\
&+&\sum_{j=0}^{\infty}\frac{1}{(2j)!}\left[\mathcal{H}_2,S\right]^{(2j)} 
\nonumber
\\
&-&i\hbar\sum_{j=0}^{\infty}\frac{1}{(2j+1)!}[\dot S, S]^{(2j)}\;,
\label{eq:off_diag}
\end{eqnarray}
\begin{eqnarray}
\mathcal{\tilde H}_{\text{diag}}&=&\sum_{j=0}^{\infty}\frac{1}{(2j)!}\left[\mathcal{H}_0+\mathcal{H}_1,S\right]^{(2j)}
\nonumber
\\
&+&\sum_{j=0}^{\infty}\frac{1}{(2j+1)!}\left[\mathcal{H}_2,S\right]^{(2j+1)}
\nonumber
\\
&-&i\hbar\sum_{j=0}^{\infty}\frac{1}{(2j+2)!}[\dot S, S]^{(2j+1)}\;.
\label{eq:diag}
\end{eqnarray}
Then, $S$ is determined by solving 
\begin{equation}
\label{eq:S}
\mathcal{\tilde H}_{\text{off-diag}}=0\;.
\end{equation}
The effective (now block-diagonal) Hamiltonian becomes $\mathcal{\tilde H}=\mathcal{\tilde H}_{\text{diag}}$.
Note that $\mathcal{\tilde H}$ as well as the term `effective Hamiltonian' is
also used to describe the block of $\mathcal{\tilde H}$ corresponding
to the relevant subpsace. 

So far no approximation has been made; now we
make use of the smallness of the perturbation. 
Following the approach of time-independent SW perturbation theory,
we aim at solving Eq. \eqref{eq:S} via
expanding $S$ as a power series in the perturbation,
\begin{eqnarray}
S=S_1+S_2+S_3+\dots\;,
\label{eq:ansatz}
\end{eqnarray}
where $S_j$ represents an operator of $j$th order in the perturbation.
Recall that in TDSW, $S$ is time dependent, and its
time derivative appears in its defining equation
\eqref{eq:S} as well as in the
effective Hamiltonian \eqref{eq:diag}.
Therefore, to separate the terms of different order in perturbing parameter in
Eq. \eqref{eq:S}, it is necessary to make an 
\emph{a priori} assumption on the 
frequency scale characterizing the magnitude of $\dot S_j$.
As the drive frequency is $\omega$, expectedly 
the frequency characterizing the time evolution of all $S_j$-s 
will be $\sim \omega$, hence we assume
$\dot S_j \sim \omega S_j$. 
In the EDSR problem defined in Sec. \ref{sec:toymodel},
the relevant subspace is the subspace of the ground-state orbital spanned by $\ket{0_\uparrow}$ and $\ket{0_\downarrow}$. Furthermore, the energy scales of the drive frequency, drive strength, Zeeman splitting
and spin-orbit coupling are much lower than the splitting between the oscillator levels $\sim\omega_0$, and all of them are treated as perturbation. This implies that $\dot S_j$ is of the order of $(j+1)$ in perturbation.

Obviously, after solving Eq. \eqref{eq:S} with this assumption,
we need to check if the obtained $S_j$ functions
are consistent with our assumption above.


From the order-by-order expansion of Eq. \eqref{eq:S},
we obtain the following hierarchy of simple algebraic equations 
for the $S_j$ matrices:
\begin{subequations}
\begin{eqnarray}
\left[\mathcal{H}_0,S_1\right]&=&-\mathcal{H}_2\;,\label{eq:s1}\\
\left[\mathcal{H}_0,S_2\right]&=&-\left[\mathcal{H}_1,S_1\right]+i\hbar \dot{S}_1\;,\label{eq:s2}\\
\left[\mathcal{H}_0,S_3\right]&=&-\left[\mathcal{H}_1,S_2\right]-\frac{1}{3}\left[\mathcal{H}_2,S_1\right]^{(2)}+i\hbar \dot{S}_2\;,\\
\left[\mathcal{H}_0,S_4\right]&=&-\left[\mathcal{H}_1,S_3\right]-\frac{1}{3}\left[\left[\mathcal{H}_2,S_1\right],S_2\right]
\nonumber \\
&&-\frac{1}{3}\left[\left[\mathcal{H}_2,S_2\right],S_1\right]+i\hbar \dot{S}_3\;,\\
\vdots\nonumber
\end{eqnarray}
\label{eq:S_eqs_2}
\end{subequations}
Once the first equation \eqref{eq:s1} is solved for $S_1(t)$,
the solution can be inserted to \eqref{eq:s1} which
then forms an algebraic equation for $S_2(t)$, etc. 
Note that since we work in the eigenbasis of $\mathcal H_0$,
the above procedure simplifies to 
subsequently solving single-variable linear equations,
which is a trivial analytical task, well suited for symbolic computation.

After obtaining the $S_j$ matrices and inserting them into 
Eq. \eqref{eq:diag}, we have an order-by-order 
expansion $\tilde {\mathcal H} = \tilde{\mathcal{H}}_{\rm diag}
=\sum_{j=0}^\infty \tilde{\mathcal H}^{(n)}$,
where 
\begin{subequations}
\begin{eqnarray}
\mathcal{\tilde H}^{(0)}&=&\mathcal{H}_0\\
\mathcal{\tilde H}^{(1)}&=&\mathcal{H}_1\\
\mathcal{\tilde H}^{(2)}&=&
	\left[\mathcal{H}_2,S_1\right]
	+\frac{1}{2}\left[\mathcal{H}_0,S_1\right]^{(2)}\\
\mathcal{\tilde H}^{(3)}&=&
	\left[\mathcal{H}_2,S_2\right]
	+\frac{1}{2}\left[\mathcal{H}_1,S_1\right]^{(2)}
	+ \frac{1}{2}\left[\left[\mathcal{H}_0,S_1\right],S_2\right]
	\nonumber	\\
	&&+\frac{1}{2}\left[\left[\mathcal{H}_0,S_2\right],S_1\right]
	-i\hbar\frac{1}{2} [\dot{S}_1,S_1]\\
\vdots\nonumber
\end{eqnarray}
\label{eq:effH_eqs}
\end{subequations}

With the use of Eqs.~\eqref{eq:S_eqs_2} we can further simplify Eqs.~\eqref{eq:effH_eqs}:
\begin{subequations}
\begin{eqnarray}
\mathcal{\tilde H}^{(0)}&=&\mathcal{H}_0\\
\mathcal{\tilde H}^{(1)}&=&\mathcal{H}_1\\
\mathcal{\tilde H}^{(2)}&=&\frac{1}{2!}\left[\mathcal{H}_2,S_1\right]\label{eq:first_corr}
\\
\mathcal{\tilde H}^{(3)}&=&\frac{1}{2!}\left[\mathcal{H}_2,S_2\right]\\
\mathcal{\tilde H}^{(4)}&=&\frac{1}{2!}\left[\mathcal{H}_2,S_3\right]-\frac{1}{4!}\left[\mathcal{H}_2,S_1\right]^{(3)}\\
\mathcal{\tilde H}^{(5)}&=&
	\frac{1}{2!}\left[\mathcal{H}_2,S_4\right]
	-\frac{1}{4!}\left[\left[\left[\mathcal{H}_2,S_1\right],S_1\right],S_2\right]
	\\
	&&-\frac{1}{4!}\left[\left[\left[\mathcal{H}_2,S_1\right],S_2\right],S_1\right]
	-\frac{1}{4!}\left[\left[\left[\mathcal{H}_2,S_2\right],S_1\right],S_1\right]
		\nonumber  \\
\vdots\nonumber
\end{eqnarray}
\label{eq:effH_eqs2}
\end{subequations}

Finally, we need to check the consistency of our assumption
for the time evolution of $S_j$ with the actual solution we obtained
for $S_j$ using that assumption. 
From \eqref{eq:s1}, $S_1$ 
inherits harmonic time-dependence from $\mathcal H_2$ 
with frequency $\omega$.
This implies that the time derivative is 
$ \dot S_1 \sim  \omega S_1$, as assumed. 
From Eq. \eqref{eq:s2}, the matrix $S_2$
might contain frequency components at $\omega$,
as well as at zero frequency and $2\omega$ (if $\mathcal H_1$ 
is time-dependent with frequency $\omega$); nevertheless, 
the $\dot S_2 \sim \omega S_2$ relation still holds, etc.

In conclusion, TDSW allows for obtaining an effective time-dependent
Hamiltonian for the relevant subspace. 
The procedure is to evaluate the transformation matrices $S_j$ up 
to the desired order via solving Eq. \eqref{eq:S_eqs_2}, 
and substituting the resulting $S_j$ matrices 
into Eq. \eqref{eq:effH_eqs2}.

\section{Rabi frequency of the fundamental resonance: relation 
to the results of Ref.~\onlinecite{Golovach-edsr}}
\label{app:gbl}

EDSR in a QD in a two-dimensional electron gas due 
to Rashba and Dresselhaus spin-orbit interactions has been described
by Golovach, Borhani and Loss (GBL) in Ref. \onlinecite{Golovach-edsr}.
Therein, the Rabi frequency of the fundamental resonance 
as a function of system parameters (magnetic field strength,
magnetic field direction, spin-orbit interaction strengths and 
ac electric field amplitude and direction) has been calculated. 
Even though the dimensionality and the spin-orbit Hamiltonian 
in the model of GBL differs from our model, 
the calculated Rabi frequencies can be compared after
a special case of the model of GBL has been reduced to 
one dimension. 
Here we show that after this dimension reduction our result for the fundamental Rabi frequency equals that of GBL.

In the model of GBL, the 2DEG lies in the $x$-$y$ plane.
We consider the special case when the confinement potential
is parabolic and has a cylindrical symmetry,
the Dresselhaus coupling vanishes, $\beta = 0$, 
the B-field is in the $y-z$ plane, and the E-field is along the $x$ axis. 
Furthermore, we project the Hamiltonian on the $y$-ground-state
orbital of the harmonic oscillator, yielding
\begin{eqnarray}
\label{eq:gblhamiltonian}
H_{\rm GBL} &=& \frac{p_x^2}{2m}
+ \frac 1 2 m \omega_0^2 x^2 + \alpha p_x \sigma_y 
+ \frac 1 2 g^* \mu_B \mathbf B \cdot \boldsymbol \sigma \nonumber \\
&+& e E_{\rm ac} x \sin \omega t
\end{eqnarray}
For simplicity, we focus on the special case 
$\mathbf{B} = (0,0,B)$ from now on.
Then, the Hamiltonian in Eq. \eqref{eq:gblhamiltonian}
is equivalent to   our Hamiltonian $\mathcal H$ at $\theta =0$.

To deduce the Rabi frequency calculated by GBL for the special case above,
 we start from their Eqs. (13) and (14), where they provide the 
time-dependent part of the effective qubit Hamiltonian
as
\begin{eqnarray}
H_{\rm GBL} = \frac 1 2 \mathbf h(t) \cdot \boldsymbol \sigma,
\end{eqnarray}
where 
\bean
\mathbf h(t) = 2 \mu_B \mathbf B \times \boldsymbol \Omega(t).
\eean
A straightforward calculation shows that 
\bean
\frac 1 2 \mathbf h (t) &=& 
\frac{\alpha e E_{\rm ac} g^* \mu_B B
}{\hbar \omega_0^2 }   \sin (\omega t) \mathbf e_x 
\\
&=&
2  
\frac{ \tilde \alpha \tilde E_{\rm ac} \tilde B
}{\hbar^2 \omega_0^2 }   \sin (\omega t) \mathbf e_x 
\eean
Note that this effective ac magnetic field is perpendicular
to the static magnetic field, which is applied in the z direction.
The Rabi frequency due to this ac magnetic field
at the fundamental resonance frequency reads
\bean
\Omega_{\rm res,GBL}^{(1)} = 
2 \frac{ \tilde \alpha \tilde E_{\rm ac} \tilde B
}{\hbar^2 \omega_0^2 },
\eean
which is identical to our result in Eq. \eqref{eq:fund_res},
if the latter is evaluated at $\theta = 0$ and 
terms above third order are dropped.


\bibliographystyle{unsrt}
\bibliographystyle{apsrev4-1}
\bibliography{TDSW}

\end{document}